\documentclass[final,nopreprintline,twocolumn,times]{elsarticle}

\usepackage{amssymb}
\usepackage{amsthm}
\usepackage{amsmath}
\usepackage[caption=false,font=normalsize,labelfont=sf,textfont=sf]{subfig}
\usepackage{wrapfig}
\usepackage{graphicx}
\usepackage{color}
\usepackage{multirow}
\usepackage[colorlinks]{hyperref}
\usepackage{bbding}
\usepackage{makecell}
\usepackage{footmisc}
\usepackage{booktabs}
\journal{}

\begin{document}

\begin{frontmatter}
	
\title{MambaEviScrib: Mamba and Evidence-Guided Consistency Enhance CNN Robustness for Scribble-Based Weakly Supervised Ultrasound Image Segmentation}

\author[shu,smart]{Xiaoxiang Han}
\ead{hanxx@shu.edu.cn}
\affiliation[shu]{organization={School of Communication and Information Engineering, Shanghai University},
	city={Shanghai},
	postcode={200444},
	country={China}}
\affiliation[smart]{organization={The SMART (Smart Medicine and AI-based Radiology Technology) Lab, Shanghai Institute for Advanced Communication and Data Science, Shanghai University},
	city={Shanghai},
	postcode={200444},
	country={China}}

\author[usst]{Xinyu Li}
\affiliation[usst]{organization={School of Health Science and Engineering, University of Shanghai for Science and Technology},
	city={Shanghai},
	postcode={200093},
	country={China}}

\author[shu,smart]{Jiang Shang}

\author[scmc]{Yiman Liu}
\affiliation[scmc]{organization={Department of Pediatric Cardiology, Shanghai Children's Medical Center, Shanghai Jiao Tong University School of Medicine},
	city={Shanghai},
	postcode={200127},
	country={China}}

\author[usst]{Keyan Chen}
\author[shu]{Shugong Xu}
 
\author[sumhs]{Qiaohong Liu\corref{cor}}
\ead{liuqh@sumhs.edu.cn}
\affiliation[sumhs]{organization={School of Medical Instruments, Shanghai University of Medicine and Health Sciences},
	city={Shanghai},
	postcode={201318},
	country={China}}

\author[shu,smart]{Qi Zhang\corref{cor}}
\ead{zhangq@t.shu.edu.cn}
\cortext[cor]{Corresponding Authors.}

\begin{abstract}
Segmenting anatomical structures and lesions from ultrasound images contributes to disease assessment, diagnosis, and treatment. Weakly supervised learning (WSL) based on sparse annotation has demonstrated the potential to reduce annotation costs. This study attempts to introduce scribble-based WSL into ultrasound image segmentation tasks. However, ultrasound images often suffer from poor contrast and unclear edges, coupled with insufficient supervision for edges, posing challenges to edge prediction. Uncertainty modeling has been proven to facilitate models in handling these issues. Nevertheless, existing uncertainty estimation paradigms lack robustness and often filter out predictions near decision boundaries, resulting in unstable edge predictions. Therefore, we propose leveraging predictions near decision boundaries effectively. Specifically, we introduce Dempster-Shafer Theory (DST) of evidence to design an Evidence-Guided Consistency (EGC) strategy. This strategy utilizes high-evidence predictions, which are more likely to occur near high-density regions, to guide the optimization of low-evidence predictions that may appear near decision boundaries. Furthermore, the varying sizes and locations of lesions in ultrasound images challenge CNNs with local receptive fields, hindering global information modeling. Therefore, we introduce Visual Mamba, a structured state space model, for long-range dependency with linear complexity, and propose a hybrid CNN-Mamba framework for both local and global information fusion. During training, the collaboration between the CNN branch and the Mamba branch draws inspiration from each other based on the EGC strategy. Extensive experiments on four ultrasound public datasets for binary-class and multi-class segmentation demonstrate the competitiveness of the proposed method. The scribble-annotated dataset and code will be made available on \url{https://github.com/GtLinyer/MambaEviScrib}.
\end{abstract}

\begin{keyword}
Weakly supervised learning, Mamba, Evidential Deep Learning, Image Segmentation, Ultrasound
\end{keyword}

\end{frontmatter}

\section{Introduction}
Medical ultrasound imaging holds a pivotal position in the field of medical diagnosis. Segmenting anatomical structures and lesions from ultrasound images plays an essential role in computer-aided diagnosis (CAD) system. It can provide valuable reference information for clinicians, such as the morphology, size, location, and relationship with surrounding tissues of organs or lesions~\cite{wang2020deep}. With the advancement of deep learning (DL)~\cite{he2016deep,vaswani2017attention,liu2023edmae}, significant progress has been made in medical image segmentation~\cite{ronneberger2015u,cao2022swin}. Generally, supervised learning methods for segmentation require large-scale pixel-wise annotated data to effectively train accurate models. However, the annotation of medical ultrasound images differs from that of natural images, requiring specialized medical expertise. Consequently, pixel-level annotation of large-scale ultrasound images is both costly and time-consuming.

\begin{figure}[!t]
	\includegraphics[width=1\linewidth]{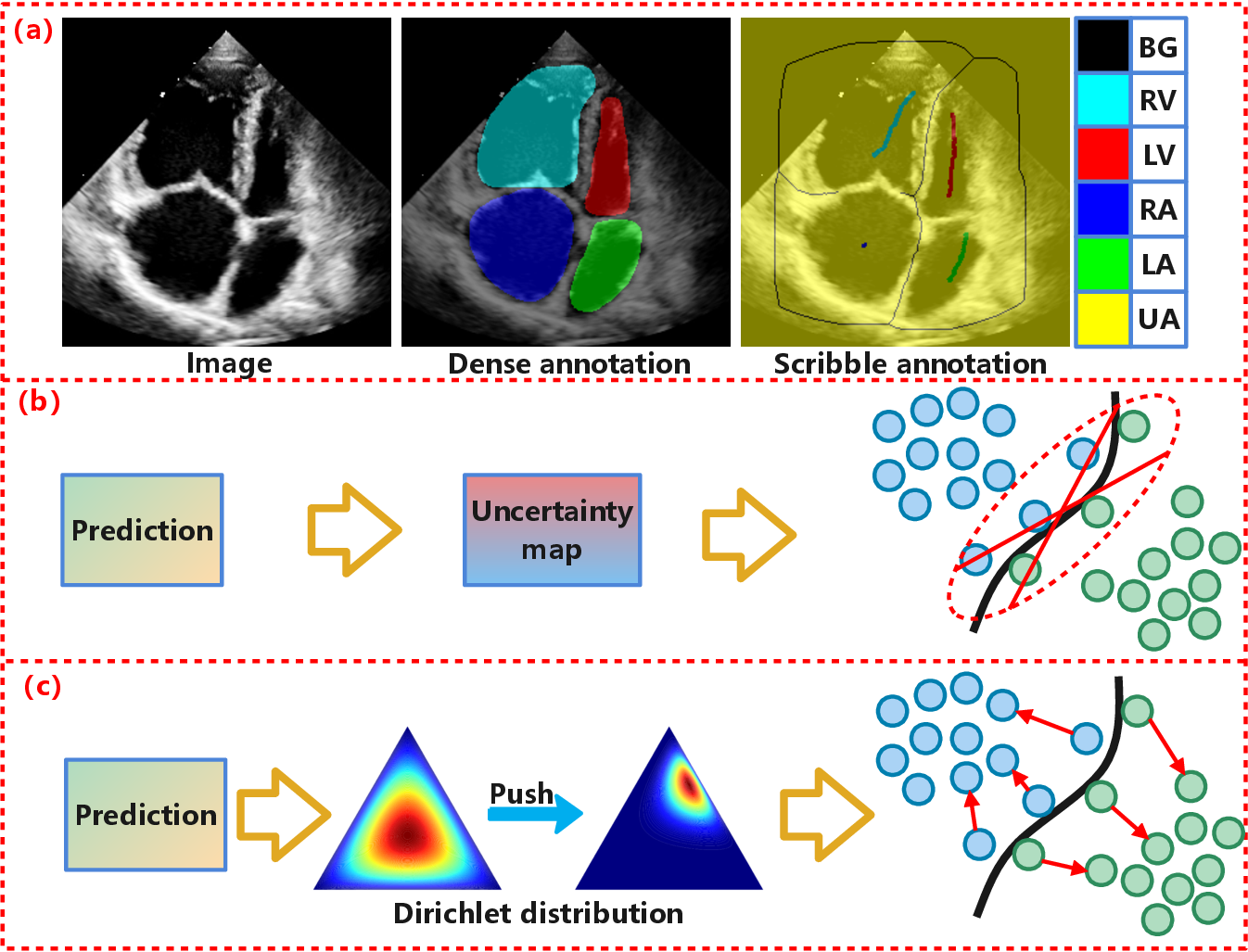}
	\caption{(a) Examples of the dense annotation and the scribble annotation. BG, RV, LV, RA, LA, and UA represent the background, right ventricle, left ventricle, right atrium, left atrium, and unannotated pixels respectively. (b) Existing uncertainty estimation methods typically discard predictions located near the decision boundary. (c) Our proposed evidence-guided consistency strategy (EGC) leverages robust evidential deep learning to guide predictions near the decision boundary towards high confidence based on evidence.}
	\label{fig:intro}
\end{figure}

To address this challenge, researchers have devoted themselves to developing DL technologies that do not rely on precise dense annotations, such as weakly supervised learning (WSL)~\cite{shen2023survey,zhao2020weakly,luo2022scribble}, semi-supervised learning~\cite{han2024deep}, and self-supervised learning~\cite{jing2020self,liu2023edmae}. This study focuses on exploring WSL method based on scribble annotations. The sparse annotation adopted by this method is easier to obtain compared to dense annotation, and offers greater convenience, versatility, and adaptability than other sparse annotation methods~\cite{tajbakhsh2020embracing}. As depicted in Figure~\ref{fig:intro} (a), by providing masks in the form of scribbles, which annotate only a small portion of pixels, this method can effectively reduce annotation costs and improve annotation efficiency. Given that a scribble annotation sparsely labels a small subset of pixels within the region of interest, the primary challenge in weakly supervised segmentation relying on scribble annotations stems from the inadequacy of training supervision information. To effectively leverage the information from unlabeled pixels, consistency regularization~\cite{zhang2022cyclemix,han2023scribble} has emerged as the most prevalent strategy. This regularization is grounded in the smoothness assumption, which posits that if two points, $x_1$ and $x_2$, are close in a high-density region, their corresponding outputs, $y_1$ and $y_2$, should also be close. Rather than directly computing distances between different input samples (which is challenging), consistency regularization synthesizes new input samples, $x'$, from the original input samples, $x$, where the proximity of $x$ and $x'$ in a semantically meaningful space is known. This approach allows the model to understand which unlabeled pixels are semantically close to labeled pixels. However, enforcing consistency in predictions across all pixels is unreasonable, as high-confidence predictions may be biased by low-confidence ones.

This prompts us to rethink the paradigm of consistency regularization. Existing methods~\cite{adiga2024anatomically} propose uncertainty estimation, then discard unreliable predictions near the decision boundary. However, considering the issues of poor contrast and unclear edges in ultrasound images, this may compromise the stability of segmentation edge predictions. Therefore, we explore how to leverage uncertainty effectively, neither enforcing uniformity in all pixel predictions nor discarding marginal decisions. Inspired by clinical consultations, clinicians with complementary domain expertise often exhibit region-specific confidence in segmentation tasks - where anatomical familiarity grants diagnostic certainty to particular areas while leaving others uncertain. During collaborative refinement, clinicians demonstrating high confidence in specific regions can guide and enhance the judgment certainty of uncertain areas through knowledge complementarity, mirroring how distinct diagnostic perspectives synergize to resolve ambiguities. Therefore, let us regard disagreements as a treasure to address the aforementioned issues. Moreover, there are two key issues to address: Firstly, how can we estimate uncertainty elegantly and robustly? Secondly, how can we construct two experts with different strengths? This is because uncertainty estimation requires evidence to support whether it stems from the model's inherent knowledge limitations. Additionally, the development of expert models with complementary strengths necessitates the integration of two architecturally distinct branch networks (e.g., those focusing on global versus local receptive fields), which is motivated by the significant heterogeneity in both dimensional scales and spatial distribution patterns exhibited by pathological lesions or target organs in ultrasound imaging.

To this end, we propose a novel scribble-based weakly supervised approach for ultrasound image segmentation, called MambaEviScrib, which comprises dual branch networks, i.e., the CNN and Mamba branches. CNN and Mamba serve as two experts with distinct advantages, where the CNN branch captures local information while the Mamba branch is responsible for capturing global features and maintaining linear computational complexity. This dual-branch architecture facilitates the fusion of global and local information. We then introduce Dempster-Shafer Theory (DST) of evidence, utilizing the Dirichlet distribution to parameterize the probability distribution of segmentation probabilities, i.e., second-order probabilities, and estimate uncertainty. An Evidence-Guided Consistency (EGC) strategy is proposed, leveraging high-evidence predictions more likely to occur near high-density regions to guide the optimization of confidence in low-evidence predictions that may appear near decision boundaries. This strategy focuses on guiding low-confidence predictions towards high confidence rather than directly enforcing consistency among divergent predictions, which contributes to the robustness and credibility of the model and enhances the stability of edge predictions. Furthermore, we design a partial evidential deep learning (pEDL) loss function for the optimization of second-order probabilities. We optimize the weakly supervised loss function, to leverage both the supervision information and the structural information to better handle edges. As illustrated in Figure~\ref{fig:intro} (b), conventional uncertainty estimation methods typically discard predictions near decision boundaries, which may lead to unstable edge predictions in ultrasound images with inherent low contrast. In contrast, our proposed evidence-guided consistency strategy [Figure~\ref{fig:intro} (c)] leverages high-evidence predictions to guide the optimization of low-evidence predictions near boundaries, enhancing edge stability through evidential reasoning.

Extensive experiments were conducted to validate MambaEviScrib on four public ultrasound datasets: CardiacUDA~\cite{yang2023graphecho}, EchoNet~\cite{ouyang2020video}, BUSI~\cite{al2020dataset} and DDTI~\cite{pedraza2015open}. Note that our approach employs only U-Net in the inference stage, and we have made no modifications to U-Net. Therefore, we enjoy a superior inference efficiency compared to other complex models.

The main contributions of this paper are as follows.
\begin{enumerate}
    \item We propose a dual-branch scribble-based weakly supervised segmentation framework comprising CNN and Mamba, which respectively extract and fuse local and global features from ultrasound images.
	
    \item The EGC strategy is developed to fully leverage predictions near decision boundaries, enhancing the stability of edge segmentation and the robustness of the model.

    \item We design a pEDL loss function, and the supervised loss functions co-optimize the generation of pseudo-labels using the features of input images and ground truth, further enhancing edge segmentation performance.
    
    \item To our best knowledge, we are the first to apply scribble-based WSL for ultrasound image segmentation. We will publicly release four ultrasound datasets along with their scribble annotations, as well as our code.
\end{enumerate}

\section{Related Works}
\subsection{DL-based ultrasound image segmentation}
With extensive research on DL in the field of medical image segmentation~\cite{ronneberger2015u,cao2022swin}, this technology has rapidly been applied to ultrasound image segmentation. However, compared with imaging techniques such as CT and MRI, ultrasound images typically have lower contrast and clarity, as well as abundant noise and artifacts, posing challenges for clinicians in analyzing and diagnosing medical conditions~\cite{ansari2024advancements}. Accurate segmentation of anatomical structures and lesions can be of great assistance. Initially, the DL-based methods primarily relied on CNN architectures. For instance, Leclerc \textit{et al.}~\cite{leclerc2019deep} evaluated that U-Net and its variants outperformed traditional methods in multi-structure ultrasound segmentation. Furthermore, some novel CNN-based network architectures~\cite{shen2020smart,xu2020convolutional} were proposed, demonstrating superior performance in ultrasound image segmentation. Liu \textit{et al.}~\cite{liu2021deep} innovatively introduced a pyramid local attention mechanism to enhance features within compact and sparse regions. The Transformer-based approach benefits from the global attention mechanism, facilitating a more comprehensive capture of contextual information within images~\cite{cao2022swin}. Li \textit{et al.}~\cite{li2023attransunet} refined the Transformer architecture, effectively reducing model complexity while enhancing segmentation accuracy, demonstrating superior performance in the segmentation of both ultrasound and pathological images. The hybrid architecture of Transformer and CNN represents a more prevalent approach for improvement, leveraging their respective strengths to achieve success in segmenting a variety of lesions from ultrasound images~\cite{chi2023hybrid,zhao2023transfsm}. For instance, Yang \textit{et al.}~\cite{yang2023cswin} devised various strategies to integrate CNN and Swin-Transformer, demonstrating a competitive performance in breast lesion segmentation on ultrasound images. The recently popular Mamba~\cite{gu2023mamba,ma2024u,ruan2024vm}, owing to its linear complexity, is poised to replace Transformer. There have been some preliminary applications of ultrasound image segmentation~\cite{ye2024p,nasiri2024vision}, for instance, Ye \textit{et al.}~\cite{ye2024hfe} introduced shape perception to apply Mmaba in segmenting the left ventricle from pediatric echocardiography. However, the aforementioned methods are all based on supervised learning, which requires large-scale dense annotations, leading to high time and labor costs. Therefore, this paper attempts to alleviate this issue in ultrasound image segmentation.

\subsection{Weakly supervised medical image segmentation}
Weakly supervised medical image segmentation utilizes weak annotations such as points~\cite{zhao2020weakly}, lines~\cite{luo2022scribble}, and bounding boxes~\cite{kulharia2020box2seg} to reduce the complexity and cost of data annotation while ensuring segmentation performance. Due to their convenience and versatility, scribble-based methods have gained popularity and achieved success in medical image segmentation~\cite{luo2022scribble}. Lin \textit{et al.}~\cite{lin2016scribblesup} conducted an early attempt by developing a graphical model that propagates information jointly from scribbles to unlabeled pixels and learns network parameters. The gated CRF loss function proposed by Obukhov \textit{et al.}~\cite{obukhov2019gated} for training unlabeled pixels aided in precise segmentation of boundaries. Lee \textit{et al.}~\cite{lee2020scribble2label} combined pseudo-labeling with label filtering to enhance the reliability of label generation. Luo \textit{et al.}~\cite{luo2022scribble} proposed a dynamic mixed pseudo-labeling method, which has achieved advantages in MR image segmentation. Furthermore, Li \textit{et al.}~\cite{li2023scribblevc} proposed ScribbleVC, which leverages vision and class embeddings via the multimodal information enhancement mechanism, unifying CNN and Transformer features for better visual feature extraction. Subsequently, they proposed ScribbleFormer, comprising three branches: a CNN branch, a Transformer branch, and an attention-guided class activation map (ACAM) branch, harnessing both local and global information~\cite{li2024scribformer}. However, most methods failed to make good use of the global information in images, and the computational cost of Transformer is relatively high. Mamba, with its distinct advantages, shows great potential.

\subsection{Uncertainty estimation and evidential deep learning}
Given the insufficiency of supervision information, the WSL method introduces uncertainty techniques to filter out unreliable predictions, thereby enhancing prediction reliability. For instance, Pan \textit{et al.}~\cite{pan2021scribble} proposed holistic operations, applying multiple manipulations to neural representations to reduce uncertainties. Liu \textit{et al.}~\cite{liu2022weakly} introduced uncertainty measurement based on Monte Carlo Dropout, enabling the calculation of consistency loss to focus solely on reliable regions. However, most methods directly discard predictions with low confidence, which may lead to inaccurate segmentation in regions such as edges. Additionally, the exclusivity of the softmax function poses difficulties in adequately describing the uncertainty within the current system.

Evidential deep learning (EDL), proposed by Sensoy \textit{et al.}~\cite{sensoy2018evidential}, aims to address the problem of out-of-distribution (OOD) samples by parameterizing the Dirichlet concentration distribution based on Dempster-Shafer theory of evidence (DST) and Subjective Logic (SL) theory~\cite{jsang2018subjective}. For controversial samples, EDL tends to provide high uncertainty rather than making incorrect predictions. Chen \textit{et al.}~\cite{chen2023evil} proposed that EVIL enhances the credibility of semi-supervised medical image segmentation based on the aforementioned theories, yet they discarded pixels with low confidence levels.

\section{Method}
\begin{figure*}[!t]
	\includegraphics[width=1\linewidth]{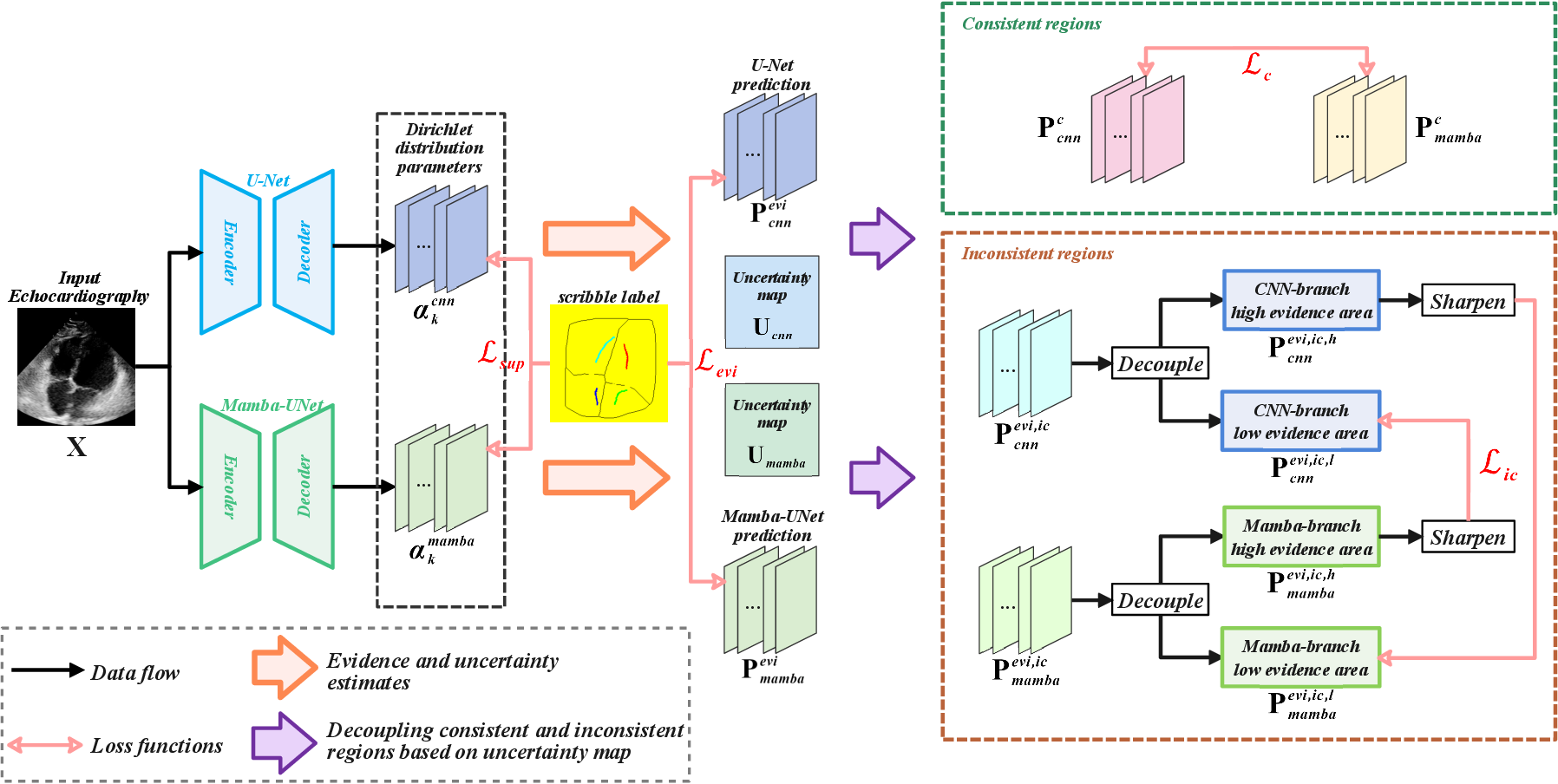}
	\caption{The pipeline of the proposed framework. The framework consists of a dual-branch network, including a CNN branch and a Mamba branch, which capture local and global features, respectively. Based on evidence theory, the uncertainty of the outputs from both networks is estimated. For ambiguous regions where the outputs of the two networks are inconsistent, predictions near decision boundaries are optimized by leveraging high-evidence predictions to guide low-evidence ones.}
	\label{fig:pipeline}
\end{figure*}

We define the 2D grayscale image of a B-mode ultrasound scan as $\mathbf{X}\in{\mathbb{R}^{W\times{H}}}$, where $W$ and $H$ represent the width and height of the image respectively. The goal of weakly supervised segmentation is to identify the class each pixel $k\in{\mathbf{X}}$ belongs to, thereby forming the semantic label map $\widehat{\mathbf{Y}}\in{\left \{ 0,1,...,K \right \}^{W\times{H}}}$, where $K=0$ represents the background class and $K>0$ indicates the target class. We utilize a dataset $\mathcal{D}=\left \{ (\mathbf{X}_i,\mathbf{Y}_i) \right \}^N_{i=1}$ containing $N$ samples for training, where $\mathbf{X}_i$ represents the input image and $\mathbf{Y}_i$ denotes the scribble annotation. In the context of multi-class segmentation, $\mathbf{Y}_i$ contains $K+1$ categories of labels: $0$ denotes unlabeled pixels, $1\sim{K-1}$ represent target pixels, and $K$ signifies background pixels. Our framework includes a U-Net and a Mamba-UNet~\cite{wang2024mamba}, and their parameters are randomized before training. Before feeding the images into the models, they undergo transformations including rotation, flipping, and color jittering. The outputs of the two networks are optimized with the EGC strategy. Furthermore, the evidence loss function and gated CRF loss function are employed to optimize the generation of pseudo labels. The details are elaborated subsequently. The proposed pipeline is illustrated in Figure~\ref{fig:pipeline}.

\subsection{CNN-Mamba dual-branch network}
Two network branches, U-Net and Mamba-UNet, are denoted as $\mathcal{F}_{cnn}(\cdot;\mathbf{\Theta}_{cnn})$ and $\mathcal{F}_{mamba}(\cdot;\mathbf{\Theta}_{mamba})$, respectively, and are highlighted in blue and green in Figure~\ref{fig:pipeline}. The input image $\mathbf{X}$ is fed into two networks separately to obtain predictions $\mathbf{P}_{cnn}, \mathbf{P}_{mamba}\in{\mathbb{R}^{K\times{W\times{H}}}}$. The whole process is expressed succinctly as:
\begin{equation}
	\mathbf{P}_{cnn} = \mathcal{F}_{cnn}(\mathbf{X};\mathbf{\Theta}_{cnn})
	\text{,}
\end{equation}
\begin{equation}
	\mathbf{P}_{mamba} = \mathcal{F}_{mamba}(\mathbf{X};\mathbf{\Theta}_{mamba})
	\text{,}
\end{equation}
where $\mathbf{\Theta}_{cnn}$ and $\mathbf{\Theta}_{mamba}$ represent the learnable parameters of the network. UNet~\cite{ronneberger2015u}, as a classic network for medical image segmentation, has been studied for a long time, offering good performance and computational efficiency. However, UNet, based on CNN, lacks the capability of capturing global information, which is enhanced under the guidance of Mamba-UNet~\cite{wang2024mamba}.  Mamba-UNet is based on Mamba, which possesses both the ability to capture global information and linear complexity.

\subsection{Evidence-guided consistency strategy}
\subsubsection{Evidence and uncertainty modeling}
\begin{figure}[ht]
	\includegraphics[width=1\linewidth]{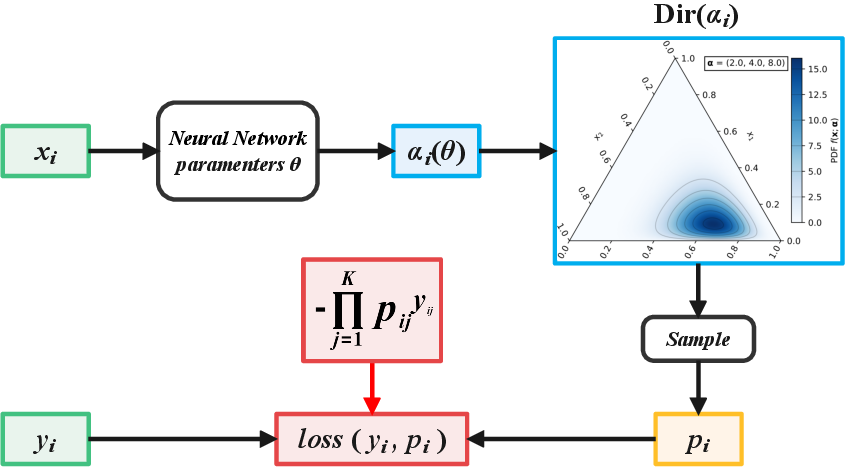}
	\caption{Evidential deep learning (EDL) pipeline and dependencies. In EDL, robust optimization of parameter $\theta$ is achieved not by directly updating parameters through loss functions defined on class probability, but rather by minimizing the conditional expectation of the loss function over the Dirichlet distribution. $x_i$ denotes the input sample, $y_i$ represents the ground truth, $\alpha_i$ signifies the parameters of the Dirichlet distribution, $Dir(\cdot)$ denotes the Dirichlet distribution, and $p_i$ indicates the class probability sampled from the Dirichlet distribution.}
	\label{fig:EDL}
\end{figure}

In traditional deep learning, the class probability $p_i$ (e.g., generated deterministically via \textit{Softmax}) is a direct function of the neural network parameters $\theta$. Thus, optimizing a loss function defined over $p_i$ via backpropagation directly updates $\theta$. However, in EDL, $p_i$ is modeled as a random variable sampled from a Dirichlet distribution $\text{Dir}(\alpha_i)$, where the concentration parameters $\alpha_i$ are generated by the neural network parameterized by $\theta$. Due to the indirect probabilistic modeling chain ($\theta \rightarrow \alpha_i \rightarrow p_i$), directly optimizing a loss function based on a single sampled instance of $p_i$ introduces significant instability from sampling noise. Thereby, EDL instead achieves robust optimization of $\theta$ by minimizing the \textit{conditional expectation} of the loss function over the Dirichlet distribution:
\begin{equation}
	\mathbb{E}_{p_i \sim \text{Dir}(\alpha_i)}  \left [ \mathcal{L}(p_i, y_i) \right ]
	\text{,}
\end{equation}
The above pipeline and dependency relationship is shown in Figure~\ref{fig:EDL}.

To model evidence and uncertainty, we introduce EDL. In EDL, each pixel is assigned belief mass and uncertainty mass. We first utilize a transformation function $f_e(\cdot)$ to derive the evidence vector $e_i$ corresponding to the $i$-th pixel ($i\in{\left \{ 0, 1, ..., W\times{H} \right \}}$). This ensures that the evidence vector remains non-negative. The process is as follows:
\begin{equation}
	\label{eq:transform_function}
	e_i = f_e \left (\mathbf{P}_i \right ) = e^{\tanh{\left ( \frac{\mathbf{P}_i}{\tau} \right ) }}
	\text{,}
\end{equation}
where $\mathbf{P}_i$ represents the output of the $i$-th pixel from the CNN or Mamba branch network, and $\tau$ denotes the scaling factor. For a segmentation task with $K$ mutually exclusive classes (including the background class), we have $e_i=\left \{ e_i^1,e_i^2,...,e_i^K \right \}$. The Dirichlet distribution with the parameters $\alpha_i$ is formulated as:
\begin{equation}
	\alpha_i^k = e_i^k + 1
	\text{,}
\end{equation}
where $k=1,2,...,K$ represents the $k$-th class. Then the $\alpha$ of the two branches is represented as $\boldsymbol{\alpha}_{cnn}$ and $\boldsymbol{\alpha}_{mamba}$. The Dirichlet strength $S_i$ is defined as:
\begin{equation}
	S_i = \sum_{k=1}^{K}{\alpha_i^k}
	\text{.}
\end{equation}
The belief mass $b_i^k$ and overall uncertainty mass $u_i$ are formulated as:
\begin{equation}
	b_i^k = \frac{e_i^k}{S_i}, \text{and } u_i = \frac{K}{S_i}
	\text{,}
\end{equation}
where $b_i^k\geq{0}$, $u_i\geq{0}$. These mass values are sum up to one, i.e.,
\begin{equation}
	u_i + \sum_{k=1}^K{b_i^k} = 1
	\text{.}
\end{equation}
This implies that uncertainty is inversely proportional to the total amount of evidence. When there is no evidence (i.e., $\sum{e_i^k}=0$), the belief of each class is zero, and the uncertainty is one. Conversely, if the total  evidence is sufficiently large, the uncertainty $u_i$ will be small, which implies that the model has high confidence in its predictions. The prediction probability distribution for the $i$-th pixel is formulated by:
\begin{equation}
	p_i^k = \frac{\alpha_i^k}{S_i}
	\text{.}
\end{equation}
Thereby, the probability distributions of $\mathbf{P}_{cnn}$, $\mathbf{P}_{mamba}$ are denoted as $\mathbf{P}_{cnn}^{evi}, \mathbf{P}_{mamba}^{evi}\in{\mathbb{R}^{K\times{W\times{H}}}}$ respectively. The Dirichlet distribution, parameterized by the evidence, represents the density of the probability distribution across the different classes predicted by the network, thereby modeling second-order probabilities and uncertainties. The Dirichlet distribution density function is formulated as:
\begin{equation}
	D \left ( \mathbf{p}_i|\boldsymbol{\alpha}_i \right ) =
	\left\{\begin{matrix}
		\frac{1}{B \left ( \boldsymbol{\alpha}_i \right )}\prod_{K}^{k=1} \left ( p_i^k \right )^{\alpha_i^k-1} & \text{for } \mathbf{p}_i\in{\mathcal{S}_K}
		\text{,}
		\\
		0 & \text{otherwise}
		\text{,}
	\end{matrix}\right.
\end{equation}
where $B(\boldsymbol{\alpha}_i)$ denotes the $K$-dimensional multinomial beta function, and $\mathcal{S}_K$ represents  the $K$-dimensional unit simplex:
\begin{equation}
	\mathcal{S}_K = \left \{ \mathbf{p} \middle| \sum_{i=1}^K{p_i = 1} \text{and } 0\leq p_1 , ... , p_K \leq 1 \right \}
	\text{.}
\end{equation}

\subsubsection{Evidence-guided consistency}
Firstly, we need to decouple the inconsistent parts predicted by the two network branches. Given that the prediction confidence and the quality of pseudo-labels are continually evolving during training, a dynamic threshold~\cite{wang2022freematch} is required to determine the regions of inconsistency. In the initial training phase, the threshold should remain as low as possible to retain more uncertain regions, promoting pseudo-label diversity and preventing the model from converging to local optima by prematurely relying on a limited set of high-confidence samples. The threshold is expressed as follows:
\begin{equation}
	\lambda_{iter} = \eta\frac{1}{B}\sum_{b=1}^B{(1-\mathbf{U})} + (1 - \eta)\lambda_{iter-1}
	\text{,}
\end{equation}
\begin{equation}
	\eta = \frac{iter}{iter_{max}}
	\text{,}
\end{equation}
where $iter$ represents the current iteration index, $\eta$ is a weight that increases with the iteration of training, $B$ denotes the batch size, and $\mathbf{U}\in{\mathbb{R}^{W\times{H}}}$ is the normalized uncertainty map. The thresholds for $\mathbf{U}_{cnn}$ and $\mathbf{U}_{mamba}$ are defined as $\lambda_{iter}^{cnn}$ and $\lambda_{iter}^{mamba}$ respectively. To promote diversity in pseudo-labels, we select the smaller one among them as the final threshold. Furthermore, The threshold is initialized to $\frac{1}{C}$ (i.e., $\lambda_0=\frac{1}{C}$), where $C$ is the number of classes. Thereby, the threshold is finally adjusted as:
\begin{equation}
	\lambda_{iter} = 
	\left\{\begin{matrix}
		\frac{1}{C} & iter=0
		\text{,}
		\\
		f_{min}\left (\lambda_{iter}^{cnn}, \lambda_{iter}^{mamba}\right ) & \text{otherwise}
		\text{,}
	\end{matrix}\right.
\end{equation}
where $f_{min}(\cdot)$ denotes the acquisition of the minimum value.

Next, we utilize the threshold to generate partition masks, which consist of mask $\mathcal{M}^c\in{ \left \{ 0,1 \right \} ^{W\times{H}}}$ for consistent regions and mask $\mathcal{M}^{ic}\in{ \left \{ 0,1 \right \} ^{W\times{H}}}$ for inconsistent regions:
\begin{equation}
	\mathcal{M}^c = \left [ \left (1-\mathbf{U}_{cnn} \right ) > \lambda_{iter} \right ] \land \left [ \left ( 1-\mathbf{U}_{mamba} \right ) > \lambda_{iter} \right ]
	\text{,}
\end{equation}
\begin{equation}
	\mathcal{M}^{ic} = \neg\mathcal{M}^c
	\text{,}
\end{equation}
where $\land$ denotes the logical AND operation, and $\neg$ represents the logical NOT operation (i.e., negation). Then, the masks are employed to partition consistent regions from the predictions. The probability distribution map of the prediction probabilities for inconsistent regions is also partitioned and regarded as evidence for the subsequent EGC strategy. The processes are expressed as follow:
\begin{equation}
	\mathbf{P}_{cnn}^c = \mathbf{P}_{cnn} \odot \mathcal{M}^{c}
	\text{,}
\end{equation}
\begin{equation}
	\mathbf{P}_{mamba}^c = \mathbf{P}_{mamba} \odot \mathcal{M}^{c}
	\text{,}
\end{equation}
\begin{equation}
	\mathbf{P}_{cnn}^{evi,ic} = \mathbf{P}_{cnn}^{evi} \odot \mathcal{M}^{ic}
	\text{,}
\end{equation}
\begin{equation}
	\mathbf{P}_{mamba}^{evi,ic} = \mathbf{P}_{mamba}^{evi} \odot \mathcal{M}^{ic}
	\text{,}
\end{equation}
where $\odot$ denotes element-wise multiplication.

\begin{figure}[ht]
	\includegraphics[width=1\linewidth]{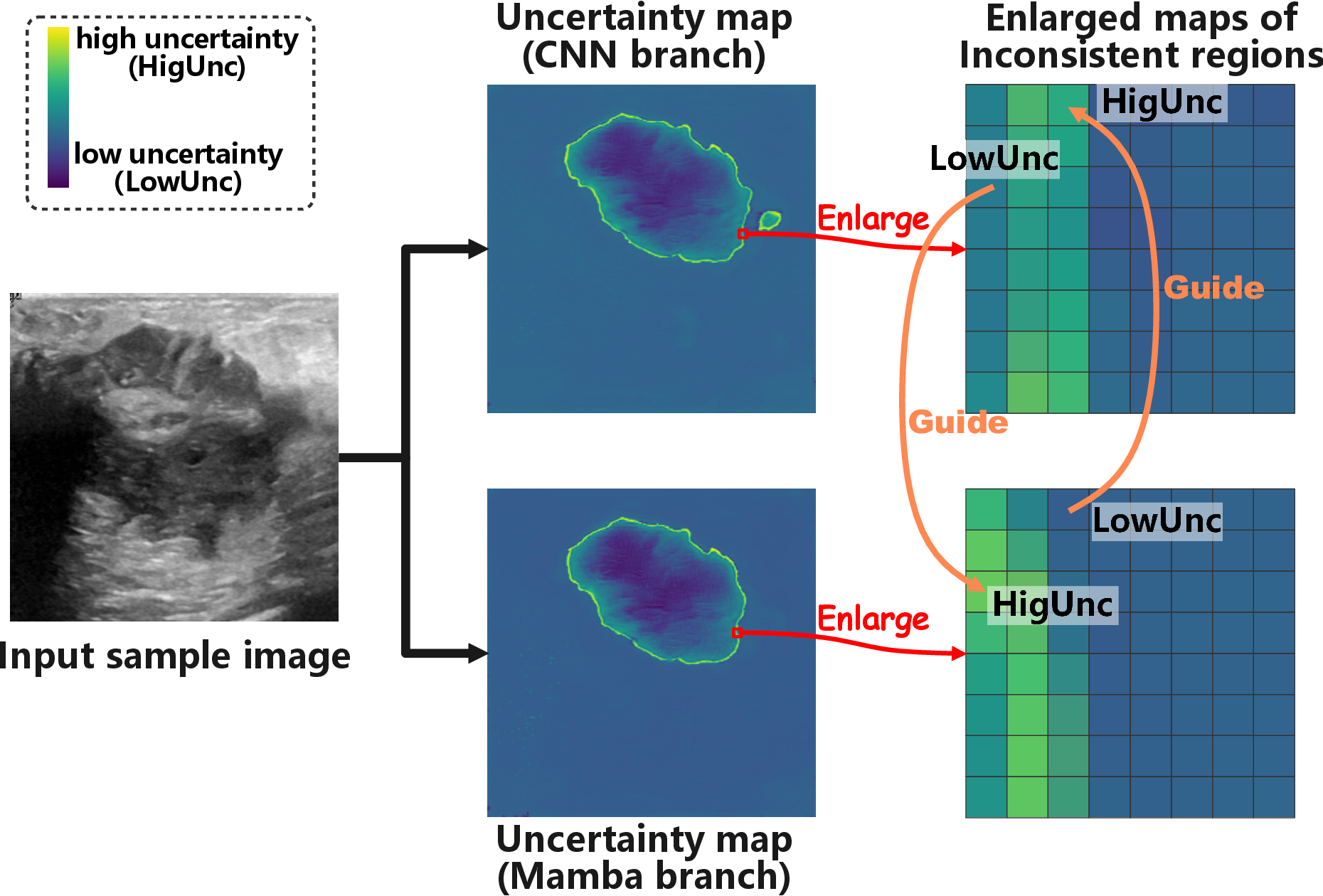}
	\caption{Schematic diagram of evidence-guided consistency strategy. For the regions with inconsistent decisions between the two network branches, we optimize the evidential confidence by leveraging the uncertainty map derived from EDL, where high evidence (low uncertainty) guides low evidence (high uncertainty).}
	\label{fig:evi_guide}
\end{figure}

Subsequently, we further decouple the inconsistent regions into predictions with low evidence that may appear near the decision boundary and predictions with high evidence that potentially occur near high-density regions. Based on the smoothness assumption, predictions from both network branches should consistently exhibit high evidence and occur near high-density regions. Consequently, we focus our efforts on optimizing probability distribution of low-evidence predictions in the vicinity of decision boundaries, as shown in Figure~\ref{fig:evi_guide}. The specific process is as follows: We first compare the evidence magnitudes between predictions from $\mathbf{P}_{cnn}^{evi,ic}$ and $\mathbf{P}_{mamba}^{evi,ic}$, identifying regions in $\mathbf{P}_{cnn}^{evi,ic}$ with higher evidence than those in $\mathbf{P}_{mamba}^{evi,ic}$. Subsequently, we sharpen the probability distribution of predictions in the high-evidence regions of $\mathbf{P}_{cnn}^{evi,ic}$ to bring their predictions closer to the high-density areas. Lastly, we utilize the predictions with higher evidence to guide those with lower evidence. The expression is presented as follows:
\begin{equation}
	\mathcal{M}_{cnn}^{h} = \mathbf{P}_{cnn}^{evi,ic} > \mathbf{P}_{mamba}^{evi,ic}
	\text{,}
\end{equation}
\begin{equation}
	\mathbf{P}_{cnn}^{evi,ic,h} = \mathbf{P}_{cnn}^{evi,ic} \odot \mathcal{M}_{cnn}^{h}
	\text{,}
\end{equation}
\begin{equation}
	\mathbf{P}_{mamba}^{evi,ic,l} = \mathbf{P}_{mamba}^{evi,ic} \odot \mathcal{M}_{cnn}^{h}
	\text{,}
\end{equation}
\begin{equation}
	\label{eq:sharping_temperature_cnn}
	\mathbf{P}_{cnn}^{evi,ic,hh} = \left ( \mathbf{P}_{cnn}^{evi,ic,h} \right ) ^{\frac{1}{\varepsilon}}
	\text{,}
\end{equation}
where $\varepsilon \in{N_{>1}}$ denotes the $\varepsilon$-th root of $\mathbf{P}_{cnn}^{evi,ic,h}$, utilized to sharpen probability distributions with high evidence. The loss function and loss calculation for guided optimization are detailed in the next subsection.

Analogously to the aforementioned process, the regions with higher evidence in $\mathbf{P}_{mamba}^{evi,ic}$ are utilized to guide the regions with lower evidence in $\mathbf{P}_{cnn}^{evi,ic}$, which can be expressed as follows:
\begin{equation}
	\mathcal{M}_{mamba}^{h} = \mathbf{P}_{mamba}^{evi,ic} > \mathbf{P}_{cnn}^{evi,ic}
	\text{,}
\end{equation}
\begin{equation}
	\mathbf{P}_{mamba}^{evi,ic,h} = \mathbf{P}_{mamba}^{evi,ic} \odot \mathcal{M}_{mamba}^{h}
	\text{,}
\end{equation}
\begin{equation}
	\mathbf{P}_{cnn}^{evi,ic,l} = \mathbf{P}_{cnn}^{evi,ic} \odot \mathcal{M}_{mamba}^{h}
	\text{,}
\end{equation}
\begin{equation}
	\label{eq:sharping_temperature_mamba}
	\mathbf{P}_{mamba}^{evi,ic,hh} = \left ( \mathbf{P}_{mamba}^{evi,ic,h} \right ) ^{\frac{1}{\varepsilon}}
	\text{.}
\end{equation}

\subsection{Loss functions}
The guidance optimization for EGC employs an L2 loss function, which is expressed as follows:
\begin{equation}
	\mathcal{L}_{EGC}^a = \ell_{l2}\left [ \mathbf{P}_{mamba}^{evi,ic,l}, \mathcal{O}_{detach}\left ( \mathbf{P}_{cnn}^{evi,ic,hh} \right ) \right ]
	\text{,}
\end{equation}
\begin{equation}
	\mathcal{L}_{EGC}^b = \ell_{l2}\left [ \mathbf{P}_{cnn}^{evi,ic,l}, \mathcal{O}_{detach}\left ( \mathbf{P}_{mamba}^{evi,ic,hh} \right ) \right ]
	\text{,}
\end{equation}
where $\mathcal{O}_{detach}(\cdot)$ denotes detachment from backpropagation, and $\ell_{l2}(\cdot, \cdot)$ represents the L2 (also known as Mean Squared Error, MSE) loss function. Thereby, the loss calculation for the inconsistent region, constrained by the EGC strategy, is as follows:
\begin{equation}
	\mathcal{L}_{ic} = \mathcal{L}_{EGC}^a + \mathcal{L}_{EGC}^b
	\text{.}
\end{equation}

For regions where the predictions of the two branch networks are consistent, we employ a cross-entropy based cross pseudo-supervision method to impose constraint. Pseudo-labels $\widehat{\mathbf{Y}}_{cnn}^{c}, \widehat{\mathbf{Y}}_{mamba}^{c}\in{\left \{ 0,1,...,K \right \}^{W\times{H}}}$ are first generated for the consistent predictions of the two branches respectively using argmax function $f_{argmax}(\cdot)$:
\begin{equation}
	\widehat{\mathbf{Y}}_{cnn}^{c} = f_{argmax} \left ( \mathbf{P}_{cnn}^c \right )
	\text{,}
\end{equation}
\begin{equation}
	\widehat{\mathbf{Y}}_{mamba}^{c} = f_{argmax} \left ( \mathbf{P}_{mamba}^c \right )
	\text{.}
\end{equation}
Then, the loss for the consistent region is calculated by the cross pseudo-supervision~\cite{chen2021semi} as follows:
\begin{equation}
	\mathcal{L}_c = \ell_{ce} \left ( \mathbf{P}_{cnn}^c, \widehat{\mathbf{Y}}_{mamba}^{c} \right ) + \ell_{ce} \left ( \mathbf{P}_{mamba}^c, \widehat{\mathbf{Y}}_{cnn}^{c} \right )
	\text{,}
\end{equation}
where $\ell_{ce}$ denotes the cross-entropy loss function.

Furthermore, for pixels annotated with scribbles, we employ partial cross-entropy (pCE) supervision on the outputs of both branches, while ignoring unlabeled pixels. And a gated conditional random field (CRF) loss is introduced to mitigate the influence of irrelevant pixels on the classification of the current pixel. This facilitates the model to better perceive morphological information within the input image, thereby emphasizing semantic boundaries. Thereby, the supervised losses calculation for the predictions from the two branch networks are as follows:
\begin{equation}
	\label{eq:loss_weight_cnn}
	\mathcal{L}_{sup}^{cnn} = \ell_{pce} \left ( \mathbf{P}_{cnn}, \mathbf{Y} \right ) + \gamma\ell_{crf} \left ( \mathbf{P}_{cnn} \right )
	\text{,}
\end{equation}
\begin{equation}
	\label{eq:loss_weight_mamba}
	\mathcal{L}_{sup}^{mamba} = \ell_{pce} \left ( \mathbf{P}_{mamba}, \mathbf{Y} \right ) +\gamma\ell_{crf} \left ( \mathbf{P}_{mamba} \right )
	\text{,}
\end{equation}
where $\ell_{pce}$ represents the partial cross-entropy (pCE) loss function, whose computational formula resembles that of the CE loss function, with the distinction being that $N$ signifies the total number of pixels labeled with scribbles, and pixels in unlabeled regions do not contribute to the calculation. $\mathbf{Y}\in{\left \{ 0,1,...,K \right \}^{W\times{H}}}$ represents the labels for scribble annotations, and $\gamma$ denotes the weight. The computational formula for the gated CRF loss function is as follows:
\begin{equation}
	\ell_{crf} = \sum_{i=1}^N{\sum_{j=1}^N{ w_{ij} \cdot \phi{ \left ( \mathbf{x}_i, \mathbf{x}_j \right )} \cdot \left ( \mathbf{p}_i - \mathbf{p}_j \right )^2}}
	\text{,}
\end{equation}
where $N$ denotes the total number of pixels, $w_{ij}$ represents the gating function to mask unexpected pixel positions, and the function $\phi(\cdot,\cdot)$ serves to quantify the similarity between pixels $\mathbf{x}_i$ and $\mathbf{x}_j$. Furthermore, $\mathbf{p}_i$ and $\mathbf{p}_j$ represent the predicted probability values for pixels $i$ and $j$, respectively.

In EDL, due to the utilization of the Dirichlet distribution to represent class probabilities, the direct optimization with the cross-entropy loss function is not feasible. Instead, the optimization target shifts to the expectation of the cross-entropy loss with respect to the Dirichlet distribution. Thereby, we introduce EDL loss function and design the partial EDL (pEDL) loss function to optimize scribble-annotated regions. The calculation of evidence loss is as follows:
\begin{equation}
	\mathcal{L}_{evi} = \ell_{pedl}\left ( \mathbf{P}_{cnn}^{evi}, \mathbf{Y} \right ) + \ell_{pedl}\left ( \mathbf{P}_{mamba}^{evi}, \mathbf{Y} \right )
	\text{,}
\end{equation}
where $\ell_{pedl}(\cdot)$ is the pEDL loss function, which comprises a expected cross-entropy (ECE) loss function and a Kullback-Leibler (KL) loss function:
\begin{equation}
	\ell_{pedl} = \frac{ f_{sum} \left [ \left (\ell_{ece} + \varphi \ell_{kl} \right ) \odot \mathcal{M}^s \right ] }{ f_{sum} \left ( \mathcal{M}^s \right ) }
	\text{,}
\end{equation}
\begin{equation}
	\varphi = f_{min} \left ( 1, \frac{2iter}{iter_{max}} \right )
	\text{,}
\end{equation}
where $\ell_{ece}(\cdot)$ is the ECE loss function, $\ell_{kl}$ is the KL loss function, $\varphi$ is an annealing coefficient, and $f_{sum}(\cdot)$ denotes summation. By gradually increasing the influence of KL divergence in the loss function through the annealing coefficient $\varphi$, the risk of misclassified samples prematurely converging to a uniform distribution is mitigated. $\mathcal{M}^s$ represents scribble-annotated regions, which are obtained by the following expression:
\begin{equation}
	\mathcal{M}^s = 0 \leq \mathbf{Y} \le K
	\text{.}
\end{equation}
Note that in this study, classes indexed as $K$ in the labels represent unannotated regions. The ECE essentially optimizes the expectation of cross-entropy under the Dirichlet distribution, necessitating the integration of cross-entropy loss over the Dirichlet distribution. Direct computation of this integral can be challenging, hence, the mean of the Dirichlet distribution is employed as the predicted probability distribution:
\begin{equation}
\begin{split}
\ell_{ece}  &= \int{\left [ \sum_{k=1}^K{-y_i^k \log{\left ( p_i^k \right )}} \right ] D \left ( \mathbf{p}_i|\boldsymbol{\alpha}_i \right )} d\mathbf{p}_i  \\  
            &= \int{\left [ \sum_{k=1}^K{-y_i^k \log{\left ( p_i^k \right )}} \right ] \frac{1}{B \left ( \boldsymbol{\alpha}_i \right )}\prod_{K}^{k=1} \left ( p_i^k \right )^{\alpha_i^k-1}} d\mathbf{p}_i \\  
            &= \sum_{k=1}^K{y_i^k \left [ \psi{ \left ( S_i \right )} - \psi{ \left ( \alpha_i^k \right )} \right]}
            \text{,}
\end{split}
\end{equation}
where $\psi{(\cdot)}$ is the \textit{digamma} function. ECE encourages the generation of evidence for positive samples across various classes. Furthermore, to reduce the evidence for negative samples, KL divergence is employed to penalize the generated evidence for negative samples:
\begin{equation}
\begin{split}
\ell_{kl}   &= KL \left [ D \left ( \mathbf{p}_i, \widetilde{\boldsymbol{\alpha}}_i \right ) \middle\| D \left ( \mathbf{p}_i, \mathbf{1} \right ) \right ] \\  
            &= \log{ \left [ \frac{\Gamma{ \left ( \sum_{k=1}^K{\widetilde{\alpha}_i^k} \right ) } }{ \Gamma{(K) } \prod_{k=1}^K{\Gamma{ \left ( \widetilde{\alpha}_i^k \right ) }}  } \right ]} \\
            &+ \sum_{k=1}^K{ \left ( \widetilde{\alpha}_i^k - 1 \right ) \left [ \psi{ \left ( \widetilde{\alpha}_i^k \right ) } - \psi{ \left ( \sum_{k=1}^K{ \widetilde{\alpha}_i^k } \right ) } \right ] }
            \text{,}
\end{split}
\end{equation}
\begin{equation}
	\widetilde{\boldsymbol{\alpha}}_i = \mathbf{y}_i + \left ( 1 - \mathbf{y}_i \right ) \odot \boldsymbol{\alpha}_i
	\text{,}
\end{equation}
where $\Gamma{(\cdot)}$ denotes the \textit{gamma} function, $D \left ( \mathbf{p}_i, \mathbf{1} \right )$ is the uniform Dirichlet distribution, $\widetilde{\boldsymbol{\alpha}}_i$ represents the Dirichlet parameter after removing misleading evidence from the predictive parameter $\boldsymbol{\alpha}_i$ of sample $i$. Therefore, the total loss is calculated as follows:
\begin{equation}
	\label{eq:all_loss}
	\mathcal{L}_{all} = \mathcal{L}_{sup}^{cnn} + \mathcal{L}_{sup}^{mamba} + \mathcal{L}_{evi} + \mathcal{L}_{ic} + \mathcal{L}_{c}
	\text{.}
\end{equation}

\section{Experiment and results}
\subsection{Datasets}
We validated the proposed method on four common public ultrasound datasets dedicated to segmentation tasks, namely CardiacUDA~\cite{yang2023graphecho}, EchoNet~\cite{ouyang2020video}, BUSI~\cite{al2020dataset}, and DDTI~\cite{pedraza2015open}, and compared it with alternative approaches. These datasets encompass diverse anatomical regions such as the breast, thyroid, and heart, encompassing both binary and multi-class segmentation tasks.

\textbf{CardiacUDA:} The CardiacUDA dataset, sourced from two anonymous hospitals, includes meticulously collected and annotated cases, approved by 5-6 experienced physicians. Each patient underwent scans in four views: parasternal long-axis left ventricle (LVLA), pulmonary artery long-axis (PALA), left ventricular short-axis (LVSA), and apical four-chamber (A4C), producing four videos per patient. Video resolutions range from $800\times600$ to $1024\times768$, depending on the scanner (Philips or Hitachi). The dataset comprises approximately 516 and 476 videos from 100 patients per hospital. Each video contains over 100 frames, covering at least one cardiac cycle, with five frames per video annotated at the pixel level for the left ventricle (LV), right ventricle (RV), left atrium (LA), and right atrium (RA). 

\textbf{EchoNet:} The EchoNet-Dynamic dataset comprises 10,030 apical four-chamber echocardiography videos sourced from clinical scans conducted at Stanford Hospital. These videos have been preprocessed to exclude non-essential content, resized to $112\times112$ pixels, and annotated with left ventricular endocardial borders at end-systole and end-diastole. We obtained these annotated frames from peers, who shared them online, and we utilized them as segmentation targets to form a new dataset for the image segmentation tasks of this study. This segmentation dataset contains 20,046 images, each accompanied by its corresponding segmentation map.

\textbf{BUSI:} Breast ultrasound images (BUSI) dataset, comprises 780 breast ultrasound images from 600 female patients aged 25 to 75, with an average image size of $500\times500$ pixels. It encompasses ultrasound images of normal, benign, and malignant breast cancer cases, along with their corresponding segmentation maps.

\textbf{DDTI:} The digital database of thyroid ultrasound images (DDTI), an open resource for the scientific community, supported by Universidad Nacional de Colombia, CIM@LAB, and Instituto de Diagnostico Medico (IDIME), encompasses 99 cases, 134 images, covering thyroiditis, cystic nodules, adenomas, and cancer. Our study utilized a preprocessed version provided by the authors of the first-place solution in the MICCAI 2020 TN-SCUI Challenge, who cleansed, cropped, and removed irrelevant regions.

\subsection{Implementation details and evaluation metrics}
For this study, images sourced from BUSI and DDTI were resized to $256\times256$ pixels for uniformity. For the BUSI dataset, we exclusively utilized samples with lesions and segmentation masks, discarding normal samples without segmentation targets. Furthermore, for samples with multiple lesion areas, the segmentation masks were merged. For the EchoNet dataset, we utilized the original size, i.e., $112\times112$ pixels. Moreover, annotated frames from CardiacUDA were extracted and resized to the same $256\times256$ pixel resolution, creating a new segmentation dataset comprising 2,250 images. To ensure rigorous evaluation, we implemented patient-wise partitioning for all four datasets during five-fold cross-validation. For CardiacUDA, we maintained strict separation of all frames from the same patient within individual folds, as the dataset contains multiple videos per patient (4 views $\times$ 100 patients). Similarly for EchoNet, we partitioned by unique patient identifiers rather than individual frames, despite its single-video-per-patient structure. This approach guarantees that no temporal neighbors from the same ultrasound sequence appear in both training and testing sets. Furthermore, our frame selection methodology for annotation explicitly avoided consecutive frames - in CardiacUDA, the five annotated frames per video were systematically spaced throughout the cardiac cycle (end-systole, end-diastole, and three intermediate phases), while EchoNet's annotations focused on distinct cardiac phases.

\begin{table*}[!t]
	\begin{center}
		\caption{Quantitative comparison results on the CardiacUDA dataset. The U-Net is trained with full supervision, serving as an upper bound. The results in bold are the best, and those in italics are the second best.}
		\label{tab:CardiacUDA}
		\setlength{\tabcolsep}{0.6mm}{
			\begin{tabular}{ c | c | l  l  l  l  l }
				\toprule
				Metrics & Methods & Left ventricle & Left atrium & Right atrium & Right ventricle & Mean \\
				\midrule
				\multirow{8}*{\makecell[c]{Dice \\ (\%) $\uparrow$}} & \makecell[c]{U-Net~\cite{ronneberger2015u} (dense label) \\ (upper bound)} & 76.49 $\pm$ 0.71\textsuperscript{\hyperlink{b}{**}} & 76.47 $\pm$ 1.08\textsuperscript{\hyperlink{a}{*}} & 79.26 $\pm$ 0.78\textsuperscript{\hyperlink{b}{**}} & 76.23 $\pm$ 0.47\textsuperscript{\hyperlink{b}{**}} & 77.11 $\pm$ 1.47\textsuperscript{\hyperlink{b}{**}} \\
				\cmidrule(lr){2-7}
				& \makecell[c]{U-Net + pCE~\cite{tang2018normalized} \\ (lower bound)} & 69.28 $\pm$ 0.87\textsuperscript{\hyperlink{b}{**}} & 51.50 $\pm$ 3.35\textsuperscript{\hyperlink{b}{**}} & 67.42 $\pm$ 2.18\textsuperscript{\hyperlink{b}{**}} & 70.83 $\pm$ 0.82\textsuperscript{\hyperlink{b}{**}} & 64.76 $\pm$ 8.18\textsuperscript{\hyperlink{b}{**}} \\
				& U-Net + Gated CRF~\cite{obukhov2019gated} & 73.10 $\pm$ 1.04\textsuperscript{\hyperlink{a}{*}} & \textit{74.06 $\pm$ 1.46}\textsuperscript{\hyperlink{a}{*}} & \textit{77.63 $\pm$ 1.19} & 73.67 $\pm$ 0.75\textsuperscript{\hyperlink{a}{*}} & \textit{74.62 $\pm$ 2.03}\textsuperscript{\hyperlink{a}{*}} \\
				& U-Net + USTM~\cite{liu2022weakly} & 68.89 $\pm$ 1.16\textsuperscript{\hyperlink{b}{**}} & 52.53 $\pm$ 0.81\textsuperscript{\hyperlink{b}{**}} & 67.83 $\pm$ 1.58\textsuperscript{\hyperlink{b}{**}} & 70.03 $\pm$ 0.97\textsuperscript{\hyperlink{b}{**}} & 64.82 $\pm$ 7.40\textsuperscript{\hyperlink{b}{**}} \\
				& DMPLS~\cite{luo2022scribble} & 66.61 $\pm$ 0.13\textsuperscript{\hyperlink{b}{**}} & 57.09 $\pm$ 1.45\textsuperscript{\hyperlink{b}{**}} & 71.59 $\pm$ 1.85\textsuperscript{\hyperlink{b}{**}} & 69.12 $\pm$ 1.81\textsuperscript{\hyperlink{b}{**}} & 66.10 $\pm$ 5.80\textsuperscript{\hyperlink{b}{**}} \\
				& ScribbleVC~\cite{li2023scribblevc} & 70.82 $\pm$ 0.87\textsuperscript{\hyperlink{b}{**}} & 69.36 $\pm$ 1.64\textsuperscript{\hyperlink{b}{**}} & 73.85 $\pm$ 0.49\textsuperscript{\hyperlink{b}{**}} & 71.79 $\pm$ 0.77\textsuperscript{\hyperlink{b}{**}} & 71.45 $\pm$ 1.92\textsuperscript{\hyperlink{b}{**}} \\
				& ScribFromer~\cite{li2024scribformer} & 62.56 $\pm$ 2.67\textsuperscript{\hyperlink{b}{**}} & 54.84 $\pm$ 2.13\textsuperscript{\hyperlink{b}{**}} & 66.76 $\pm$ 1.34\textsuperscript{\hyperlink{b}{**}} & 63.44 $\pm$ 2.85\textsuperscript{\hyperlink{b}{**}} & 61.90 $\pm$ 4.96\textsuperscript{\hyperlink{b}{**}} \\
				& DMSPS~\cite{han2024dmsps} & 73.41 $\pm$ 0.79\textsuperscript{\hyperlink{a}{*}} & 70.11 $\pm$ 1.00\textsuperscript{\hyperlink{b}{**}} & 75.71 $\pm$ 0.58\textsuperscript{\hyperlink{b}{**}} & \textit{73.88 $\pm$ 0.82} & 73.28 $\pm$ 2.20\textsuperscript{\hyperlink{b}{**}} \\
				& Bayesian\_WSS~\cite{zheng2024bayesian} & \textit{73.79 $\pm$ 0.94} & 73.59 $\pm$ 1.60 & 76.52 $\pm$ 0.52\textsuperscript{\hyperlink{a}{*}} & 73.48 $\pm$ 1.01 & 74.34 $\pm$ 1.63\textsuperscript{\hyperlink{a}{*}} \\
				& Ours & \textbf{74.36 $\pm$ 0.44} & \textbf{75.16 $\pm$ 1.07} & \textbf{77.85 $\pm$ 0.77} & \textbf{74.19 $\pm$ 0.79} & \textbf{75.39 $\pm$ 1.67} \\
				\midrule
				\multirow{8}*{\makecell[c]{95HD \\ (pixel) $\downarrow$}} & \makecell[c]{U-Net~\cite{ronneberger2015u} (dense label) \\ (upper bound)} & 16.25 $\pm$ 0.69\textsuperscript{\hyperlink{b}{**}} & 8.94 $\pm$ 0.92\textsuperscript{\hyperlink{b}{**}} & 6.43 $\pm$ 0.28\textsuperscript{\hyperlink{b}{**}} & 7.59 $\pm$ 0.43\textsuperscript{\hyperlink{b}{**}} & 9.80 $\pm$ 3.97\textsuperscript{\hyperlink{b}{**}} \\
				\cmidrule(lr){2-7}
				& \makecell[c]{U-Net + pCE~\cite{tang2018normalized} \\ (lower bound)} & 25.03 $\pm$ 2.25\textsuperscript{\hyperlink{b}{**}} & 80.88 $\pm$ 5.41\textsuperscript{\hyperlink{b}{**}} & 28.82 $\pm$ 7.61\textsuperscript{\hyperlink{b}{**}} & 15.63 $\pm$ 0.49\textsuperscript{\hyperlink{b}{**}} & 37.59 $\pm$ 26.48\textsuperscript{\hyperlink{b}{**}} \\
				& U-Net + Gated CRF~\cite{obukhov2019gated} & 18.72 $\pm$ 0.58 & 11.43 $\pm$ 1.34\textsuperscript{\hyperlink{a}{*}} & \textit{7.68 $\pm$ 0.35} & 9.84 $\pm$ 0.24\textsuperscript{\hyperlink{a}{*}} & \textit{11.92 $\pm$ 4.43} \\
				& U-Net + USTM~\cite{liu2022weakly} & 27.08 $\pm$ 2.39\textsuperscript{\hyperlink{b}{**}} & 76.57 $\pm$ 2.63\textsuperscript{\hyperlink{b}{**}} & 37.64 $\pm$ 11.68\textsuperscript{\hyperlink{b}{**}} & 13.66 $\pm$ 1.27\textsuperscript{\hyperlink{b}{**}} & 38.74 $\pm$ 24.70\textsuperscript{\hyperlink{b}{**}} \\
				& DMPLS~\cite{luo2022scribble} & 26.65 $\pm$ 0.66\textsuperscript{\hyperlink{b}{**}} & 59.85 $\pm$ 2.52\textsuperscript{\hyperlink{b}{**}} & 14.79 $\pm$ 2.46\textsuperscript{\hyperlink{b}{**}} & 13.86 $\pm$ 0.75\textsuperscript{\hyperlink{b}{**}} & 28.79 $\pm$ 19.19\textsuperscript{\hyperlink{b}{**}} \\
				& ScribbleVC~\cite{li2023scribblevc} & 20.19 $\pm$ 0.56\textsuperscript{\hyperlink{b}{**}} & 15.12 $\pm$ 1.41\textsuperscript{\hyperlink{b}{**}} & 10.98 $\pm$ 0.60\textsuperscript{\hyperlink{b}{**}} & 10.29 $\pm$ 0.26\textsuperscript{\hyperlink{b}{**}} & 14.14 $\pm$ 4.12\textsuperscript{\hyperlink{b}{**}} \\
				& ScribFromer~\cite{li2024scribformer} & 32.74 $\pm$ 5.09\textsuperscript{\hyperlink{b}{**}} & 38.68 $\pm$ 14.59\textsuperscript{\hyperlink{a}{*}} & 24.34 $\pm$ 6.65\textsuperscript{\hyperlink{b}{**}} & 21.86 $\pm$ 2.53\textsuperscript{\hyperlink{b}{**}} & 29.41 $\pm$ 10.40\textsuperscript{\hyperlink{b}{**}} \\
				& DMSPS~\cite{han2024dmsps} & 18.81 $\pm$ 0.75 & 13.69 $\pm$ 0.53\textsuperscript{\hyperlink{b}{**}} & 8.49 $\pm$ 0.32\textsuperscript{\hyperlink{b}{**}} & \textit{9.36 $\pm$ 0.46} & 12.59 $\pm$ 4.23\textsuperscript{\hyperlink{a}{*}} \\
				& Bayesian\_WSS~\cite{zheng2024bayesian} & \textit{18.60 $\pm$ 0.86} & \textit{11.25 $\pm$ 1.07} & 7.96 $\pm$ 0.81 & 10.49 $\pm$ 0.51\textsuperscript{\hyperlink{b}{**}} & 12.08 $\pm$ 4.14\textsuperscript{\hyperlink{a}{*}} \\
				& Ours & \textbf{18.57 $\pm$ 0.39} & \textbf{10.56 $\pm$ 1.41} & \textbf{7.64 $\pm$ 0.37} & \textbf{9.64 $\pm$ 0.12} & \textbf{11.60 $\pm$ 4.33} \\
				\bottomrule
				\multicolumn{7}{l}{\hypertarget{a}{\textcolor{blue}{*}} The result is significantly different from ours with $p < 0.05$ via paired t-test. \hypertarget{b}{\textcolor{blue}{**}} $p < 0.01$.}
		\end{tabular}}
	\end{center}
\end{table*}

\begin{figure*}[ht]
	\includegraphics[width=1\linewidth]{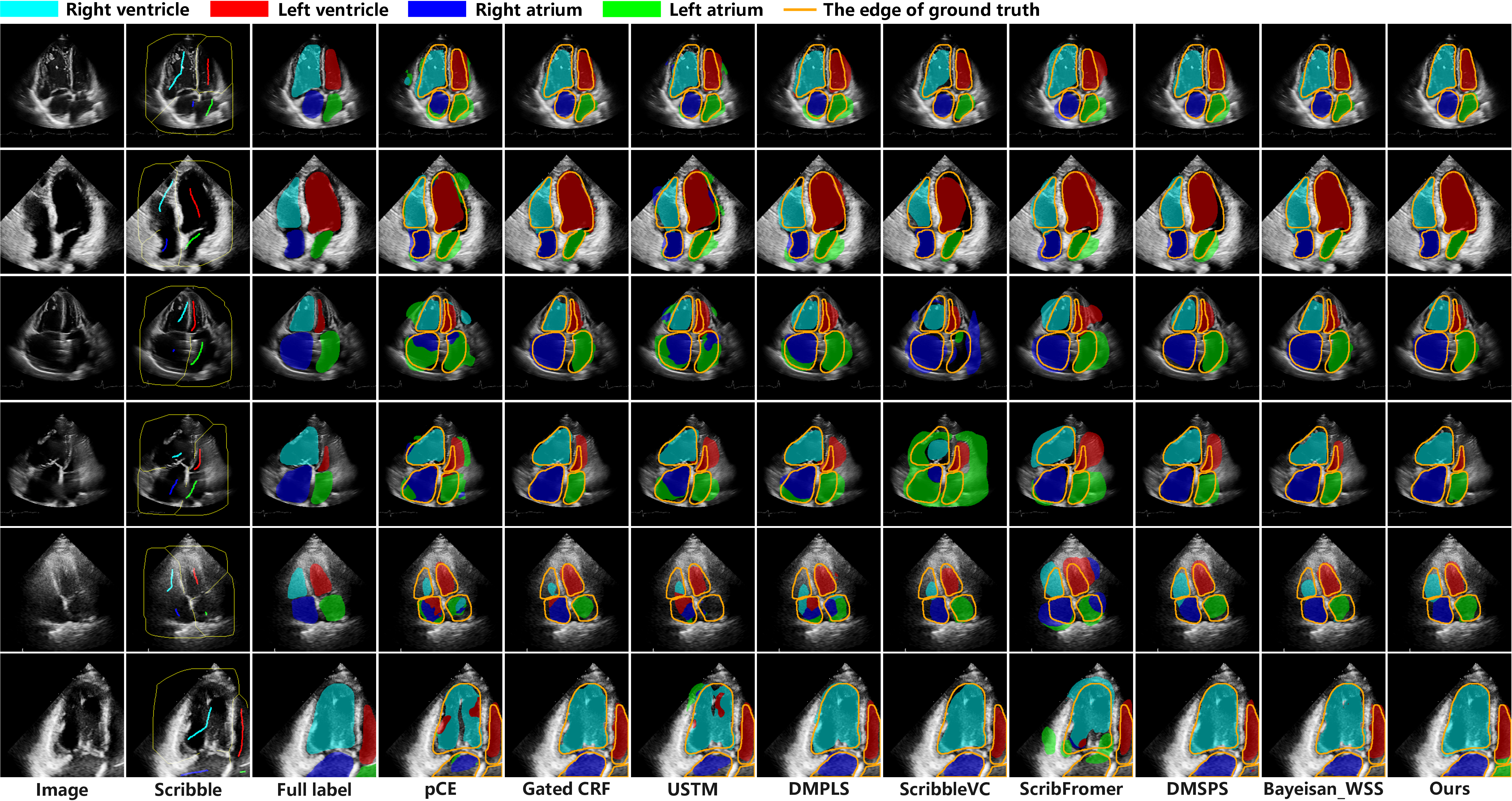}
	\caption{Visualization of experimental results comparison on the CardiacUDA dataset. The \textit{Scribble} column represents scribble annotations, while the \textit{Full label} column indicates full dense annotations. The yellow lines indicate the ground truth.}
	\label{fig:CardiacUDA}
\end{figure*}

All scribble annotations were generated by the WSL4MIS code\footnote{\url{https://github.com/HiLab-git/WSL4MIS} \label{fn:wsl4mis}} provided by HiLab at UESTC. This method initially extracts the two largest connected components from a binary mask. It then detects and processes branching structures in the image, removing unnecessary branches. Finally, it generates a skeletonization representation of the 2D image, which is used to create doodles. In the generated annotations, the background class is represented as $0$, and the unlabeled class is denoted as $K$ (i.e., the largest number excluding other categories). The scribble annotations we generated are relatively sparse. We evaluated the proposed method and other methods on all datasets using five-fold cross-validation. The optimizer used to train the model is Stochastic Gradient Descent (SGD), with a weight decay of $10^{-4}$ and a momentum of 0.9, to minimize the joint objective function Eq.~\ref{eq:all_loss}. We implemented our proposed method and other comparison methods based on Python 3.8, PyTorch 1.12, and the WSL4MIS codebase\footref{fn:wsl4mis}, and trained them on a server with 2 Nvidia Geforce RTX 3080 GPUs, totaling 20GB of memory. Images were preprocessed before training, including random horizontal or vertical flipping, random angle rotation, random equalization, and random adjustments of brightness, contrast, saturation, and hue. The learning rate was adjusted online by the poly learning rate scheduler~\cite{luo2022scribble}:
\begin{equation}
	LR_{iter} = \left ( 1.0 - \frac{iter}{iter_{max}} \right )^{0.9}LR_0
	\text{,}
\end{equation}
where $LR_{iter}$ denotes the learning rate at the $iter$-th iteration. The batch size and total iterations ($iter$) were set to 12 and 60k, respectively. Hyper-parameters $\tau$. $\epsilon$ and $\gamma$ were set to 0.25, 0.5 and 0.1, respectively. For a \textbf{fair} comparison, we employed the output of U-Net as the final result in the testing phase without applying any post-processing methods. And all experiments were conducted in the same experimental environment.

We quantitatively evaluated all methods using 4 metrics widely adopted in weakly or semi-supervised medical image segmentation tasks. The evaluation metrics encompass the Dice score, Jaccard index, average surface distance (ASD), and 95\% Hausdorff distance (95HD). The Dice score and Jaccard index assess pixel-level overlap between the ground truth and predictions, while ASD and 95HD measure the surface distances between them. Note that for the multi-class segmentation task on the CardiacUDA dataset, we only present Dice and 95HD as evaluation metrics.

In our experiments, the mean and standard deviation values were calculated based on the performance metrics (e.g., Dice score, Jaccard index, 95HD, ASD) obtained from five-fold cross-validation. For each fold, the model was trained and tested on distinct subsets of the dataset, and the performance metrics were recorded. The mean value represents the average performance across all folds, while the standard deviation quantifies the variability in the results. Specifically, the mean ($\mu$) and standard deviation ($\sigma$) were computed as follows:
\begin{equation}
	\mu = \frac{1}{N} \sum_{i=1}^{N}{x_i}
	\text{ , and }
	\sigma = \sqrt{\frac{1}{N} \sum_{i=1}^{N}{\left ( x_i - \mu\right )^2}}
	\text{,}
\end{equation}
where $x_i$ denotes the performance metric for the $i$-th fold, and $N=5$ is the number of folds.

\begin{table*}[ht]
	\begin{center}
		\caption{Quantitative comparison results on the EchoNet dataset. The U-Net is trained with full supervision, serving as an upper bound. The results in bold are the best, and those in italics are the second best.}
		\label{tab:EchoNet}
		\setlength{\tabcolsep}{3mm}{
			\begin{tabular}{ c | l  l  l  l }
				\toprule
				Methods & Dice (\%) $\uparrow$ & Jaccard (\%) $\uparrow$ & 95HD (pixel) $\downarrow$ & ASD (pixel) $\downarrow$ \\
				\midrule
				U-Net ~\cite{ronneberger2015u} (upper bound) & 92.64 $\pm$ 0.08\textsuperscript{\hyperlink{b}{**}} & 86.57 $\pm$ 0.10\textsuperscript{\hyperlink{b}{**}} & 2.79 $\pm$ 0.06\textsuperscript{\hyperlink{b}{**}} & 1.12 $\pm$ 0.02\textsuperscript{\hyperlink{b}{**}} \\
				\midrule
				U-Net + pCE~\cite{tang2018normalized} (lower bound) & 68.57 $\pm$ 0.81\textsuperscript{\hyperlink{b}{**}} & 53.53 $\pm$ 0.97\textsuperscript{\hyperlink{b}{**}} & 12.38 $\pm$ 0.68\textsuperscript{\hyperlink{b}{**}} & 6.38 $\pm$ 0.17\textsuperscript{\hyperlink{b}{**}} \\
				U-Net + Gated CRF~\cite{obukhov2019gated} & 84.31 $\pm$ 0.59 & 73.60 $\pm$ 0.84 & 6.14 $\pm$ 0.23 & 2.58 $\pm$ 0.13\textsuperscript{\hyperlink{a}{*}} \\
				U-Net + USTM~\cite{liu2022weakly} & 71.60 $\pm$ 3.57\textsuperscript{\hyperlink{b}{**}} & 57.19 $\pm$ 4.19\textsuperscript{\hyperlink{b}{**}} & 12.35 $\pm$ 1.48\textsuperscript{\hyperlink{b}{**}} & 5.51 $\pm$ 1.06\textsuperscript{\hyperlink{b}{**}} \\
				DMPLS~\cite{luo2022scribble} & 74.52 $\pm$ 1.05\textsuperscript{\hyperlink{b}{**}} & 60.54 $\pm$ 1.41\textsuperscript{\hyperlink{b}{**}} & 10.30 $\pm$ 0.63\textsuperscript{\hyperlink{b}{**}} & 4.00 $\pm$ 0.52\textsuperscript{\hyperlink{b}{**}} \\
				ScribbleVC~\cite{li2023scribblevc} & \textit{84.77 $\pm$ 0.58} & \textit{74.19 $\pm$ 0.82} & \textit{6.01 $\pm$ 0.18} & \textit{2.43 $\pm$ 0.10} \\
				ScribFromer~\cite{li2024scribformer} & 63.53 $\pm$ 1.71\textsuperscript{\hyperlink{b}{**}} & 47.69 $\pm$ 1.74\textsuperscript{\hyperlink{b}{**}} & 15.48 $\pm$ 0.70\textsuperscript{\hyperlink{b}{**}} & 7.90 $\pm$ 0.58\textsuperscript{\hyperlink{b}{**}} \\
				DMSPS~\cite{han2024dmsps} & 80.70 $\pm$ 1.60\textsuperscript{\hyperlink{b}{**}} & 68.66 $\pm$ 2.14\textsuperscript{\hyperlink{b}{**}} & 7.92 $\pm$ 0.75\textsuperscript{\hyperlink{b}{**}} & 3.26 $\pm$ 0.40\textsuperscript{\hyperlink{a}{*}} \\
				Bayesian\_WSS~\cite{zheng2024bayesian} & 78.63 $\pm$ 0.14\textsuperscript{\hyperlink{b}{**}} & 65.55 $\pm$ 0.21\textsuperscript{\hyperlink{b}{**}} & 9.48 $\pm$ 0.15\textsuperscript{\hyperlink{b}{**}} & 3.89 $\pm$ 0.03\textsuperscript{\hyperlink{b}{**}} \\
				Ours & \textbf{85.10 $\pm$ 0.52} & \textbf{74.78 $\pm$ 0.73} & \textbf{5.95 $\pm$ 0.24} & \textbf{2.38 $\pm$ 0.11} \\
				\bottomrule
				\multicolumn{5}{l}{\hypertarget{a}{\textcolor{blue}{*}} The result is significantly different from ours with $p < 0.05$ via paired t-test. \hypertarget{b}{\textcolor{blue}{**}} $p < 0.01$.}
		\end{tabular}}
	\end{center}
\end{table*}

\begin{figure*}[!t]
\includegraphics[width=1\linewidth]{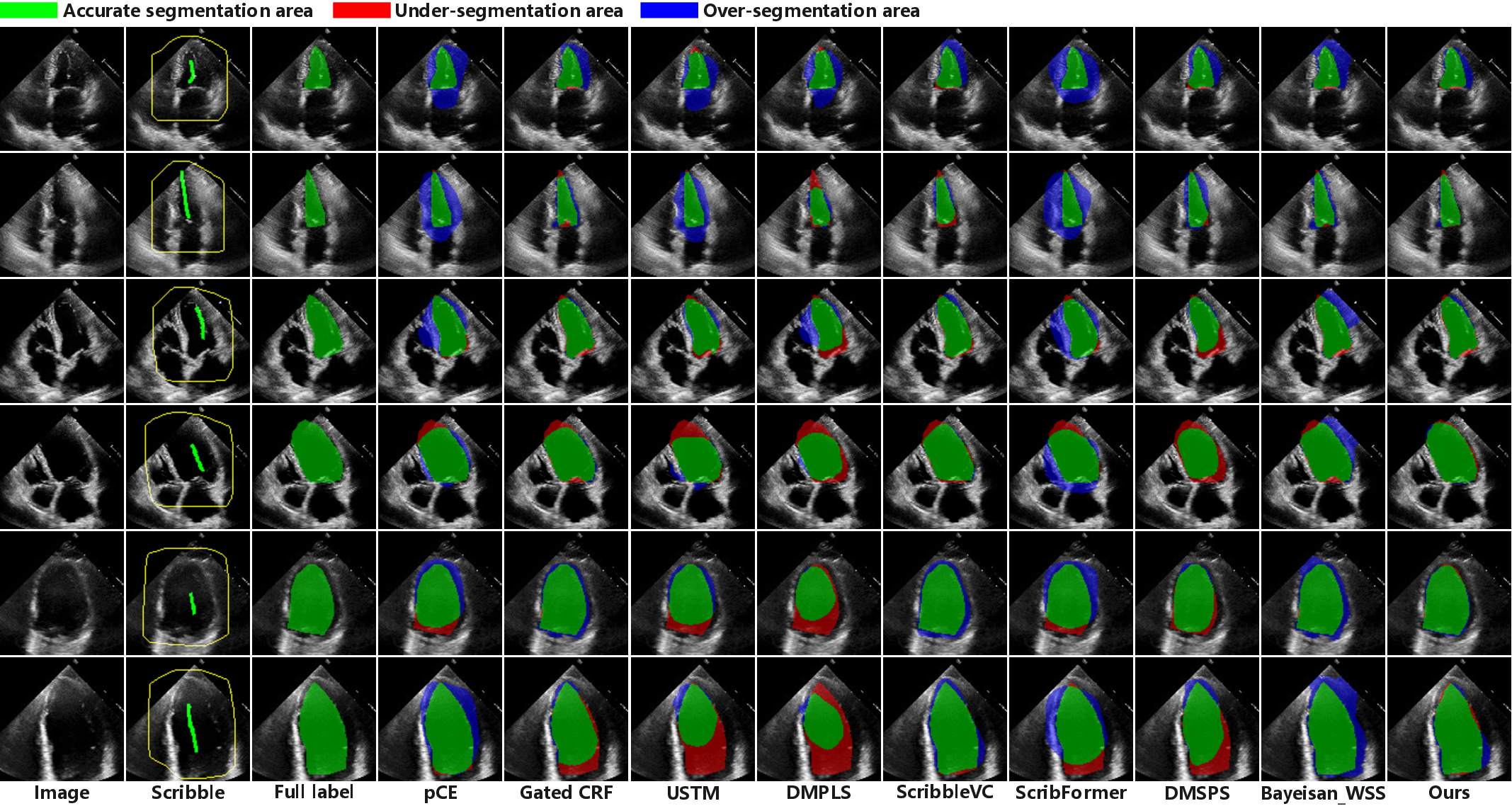}
\caption{Visualization of comparison experimental results on the EchoNet dataset. The \textit{Scribble} column represents scribble annotations, while the \textit{Full label} column indicates full dense annotations. In scribble annotations, the yellow area represents unlabeled area. In full labels, the green area denotes the ground truth. In other segmentation results, green area indicates accurate segmentation, i.e., the overlap between the segmentation result and the ground truth. Red area represents under-segmentation (false negatives), while blue area indicate over-segmentation (false positives).}
\label{fig:EchoNet}
\end{figure*}

\subsection{Comparison with existing methods}

\begin{table*}[ht]
	\begin{center}
		\caption{Quantitative comparison results on the BUSI dataset. The U-Net is trained with full supervision, serving as an upper bound. The results in bold are the best, and those in italics are the second best.}
		\label{tab:BUSI}
		\setlength{\tabcolsep}{3mm}{
			\begin{tabular}{ c | l  l  l  l }
				\toprule
				Methods & Dice (\%) $\uparrow$ & Jaccard (\%) $\uparrow$ & 95HD (pixel) $\downarrow$ & ASD (pixel) $\downarrow$ \\
				\midrule
				U-Net ~\cite{ronneberger2015u} (upper bound) & 77.91 $\pm$ 2.54\textsuperscript{\hyperlink{b}{**}} & 69.52 $\pm$ 2.45\textsuperscript{\hyperlink{b}{**}} & 29.75 $\pm$ 3.86 & 11.18 $\pm$ 1.76\\
				\midrule
				U-Net + pCE~\cite{tang2018normalized} (lower bound) & 62.89 $\pm$ 2.90\textsuperscript{\hyperlink{b}{**}} & 50.36 $\pm$ 3.39\textsuperscript{\hyperlink{b}{**}} & 47.03 $\pm$ 5.77\textsuperscript{\hyperlink{b}{**}} & 18.96 $\pm$ 2.27\textsuperscript{\hyperlink{b}{**}} \\
				U-Net + Gated CRF~\cite{obukhov2019gated} & 71.39 $\pm$ 2.66\textsuperscript{\hyperlink{a}{*}} & \textit{60.84 $\pm$ 2.37}\textsuperscript{\hyperlink{b}{**}} & 35.90 $\pm$ 2.86\textsuperscript{\hyperlink{a}{*}} & 13.54 $\pm$ 1.49\textsuperscript{\hyperlink{a}{*}} \\
				U-Net + USTM~\cite{liu2022weakly} & 61.99 $\pm$ 3.05\textsuperscript{\hyperlink{b}{**}} & 49.11 $\pm$ 3.05\textsuperscript{\hyperlink{b}{**}} & 47.50 $\pm$ 5.16\textsuperscript{\hyperlink{b}{**}} & 19.20 $\pm$ 2.41\textsuperscript{\hyperlink{b}{**}} \\
				DMPLS~\cite{luo2022scribble} & 65.09 $\pm$ 3.44\textsuperscript{\hyperlink{b}{**}} & 52.85 $\pm$ 3.45\textsuperscript{\hyperlink{b}{**}} & 36.27 $\pm$ 3.21\textsuperscript{\hyperlink{b}{**}} & 14.63 $\pm$ 1.58\textsuperscript{\hyperlink{b}{**}} \\
				ScribbleVC~\cite{li2023scribblevc} & 69.92 $\pm$ 3.87\textsuperscript{\hyperlink{b}{**}} & 57.90 $\pm$ 3.90\textsuperscript{\hyperlink{b}{**}} & \textit{31.23 $\pm$ 5.67} & 12.55 $\pm$ 2.85 \\
				ScribFromer~\cite{li2024scribformer} & 65.09 $\pm$ 3.44\textsuperscript{\hyperlink{b}{**}} & 52.85 $\pm$ 3.45\textsuperscript{\hyperlink{b}{**}} & 36.27 $\pm$ 3.21\textsuperscript{\hyperlink{b}{**}} & 14.63 $\pm$ 1.58\textsuperscript{\hyperlink{b}{**}} \\
				DMSPS~\cite{han2024dmsps} & \textit{71.75 $\pm$ 2.14}\textsuperscript{\hyperlink{a}{*}} & 60.66 $\pm$ 1.93\textsuperscript{\hyperlink{a}{*}} & 31.26 $\pm$ 3.40 & \textit{12.18 $\pm$ 1.46} \\
				Bayesian\_WSS~\cite{zheng2024bayesian} & 68.42 $\pm$ 3.02\textsuperscript{\hyperlink{a}{*}} & 57.93 $\pm$ 3.25\textsuperscript{\hyperlink{a}{*}} & 34.78 $\pm$ 8.13 & 13.30 $\pm$ 3.35 \\
				Ours & \textbf{74.07 $\pm$ 3.27} & \textbf{63.99 $\pm$ 3.18} & \textbf{31.10 $\pm$ 2.73} & \textbf{11.71 $\pm$ 1.60} \\
				\bottomrule
				\multicolumn{5}{l}{\hypertarget{a}{\textcolor{blue}{*}} The result is significantly different from ours with $p < 0.05$ via paired t-test. \hypertarget{b}{\textcolor{blue}{**}} $p < 0.01$.}
		\end{tabular}}
	\end{center}
\end{table*}

\begin{figure*}[!t]
	\includegraphics[width=1\linewidth]{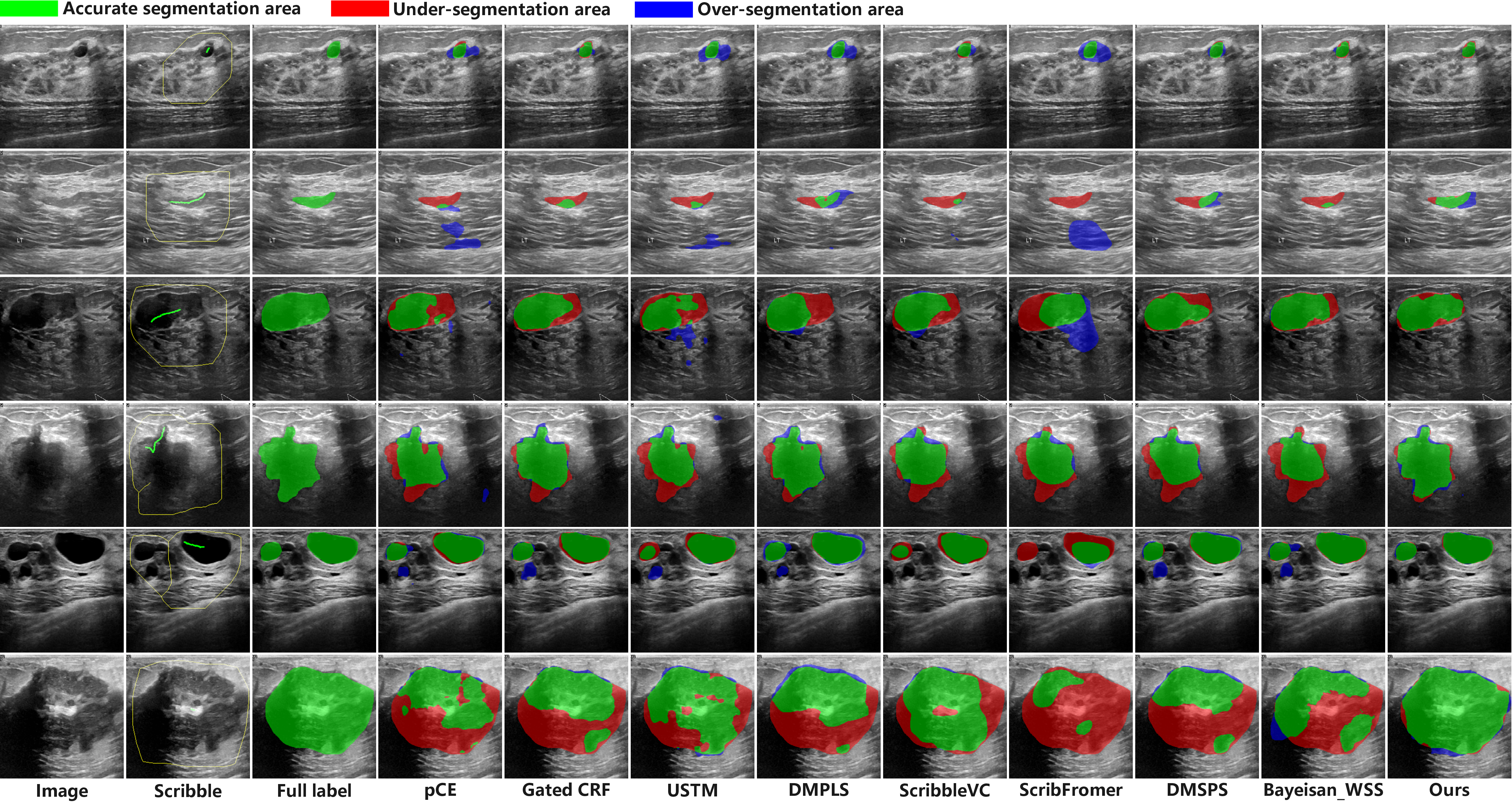}
	\caption{Visualization of comparison experimental results on the BUSI dataset. The \textit{Scribble} column represents scribble annotations, while the \textit{Full label} column indicates full dense annotations. In scribble annotations, the yellow area represents unlabeled area. In full labels, the green area denotes the ground truth. In other segmentation results, green area indicates accurate segmentation, i.e., the overlap between the segmentation result and the ground truth. Red area represents under-segmentation (false negatives), while blue area indicate over-segmentation (false positives).}
	\label{fig:BUSI}
\end{figure*}

\begin{table*}[ht]
\begin{center}
	\caption{Quantitative comparison results on the DDTI dataset. The U-Net is trained with full supervision, serving as an upper bound. The results in bold are the best, and those in italics are the second best.}
	\label{tab:DDTI}
	\setlength{\tabcolsep}{3mm}{
		\begin{tabular}{ c | l  l  l  l }
			\toprule
			Methods & Dice (\%) $\uparrow$ & Jaccard (\%) $\uparrow$ & 95HD (pixel) $\downarrow$ & ASD (pixel) $\downarrow$ \\
			\midrule
			U-Net ~\cite{ronneberger2015u} (upper bound) & 79.23 $\pm$ 1.74\textsuperscript{\hyperlink{b}{**}} & 68.57 $\pm$ 1.66\textsuperscript{\hyperlink{b}{**}} & 27.03 $\pm$ 1.69\textsuperscript{\hyperlink{a}{*}} & 10.15 $\pm$ 0.85\textsuperscript{\hyperlink{a}{*}} \\
			\midrule
			U-Net + pCE~\cite{tang2018normalized} (lower bound) & 64.27 $\pm$ 1.53\textsuperscript{\hyperlink{a}{*}} & 50.22 $\pm$ 1.45 & 39.53 $\pm$ 1.89\textsuperscript{\hyperlink{a}{*}} & 19.00 $\pm$ 1.14\textsuperscript{\hyperlink{a}{*}} \\
			U-Net + Gated CRF~\cite{obukhov2019gated} & \textit{66.53 $\pm$ 1.02} & \textit{52.48 $\pm$ 1.12} & \textit{35.58 $\pm$ 0.96} & \textit{16.12 $\pm$ 0.69} \\
			U-Net + USTM~\cite{liu2022weakly} & 63.86 $\pm$ 1.66 & 49.65 $\pm$ 1.58 & 39.61 $\pm$ 2.15\textsuperscript{\hyperlink{a}{*}} & 19.08 $\pm$ 1.47\textsuperscript{\hyperlink{a}{*}} \\
			DMPLS~\cite{luo2022scribble} & 62.88 $\pm$ 1.59\textsuperscript{\hyperlink{a}{*}} & 48.71 $\pm$ 1.54\textsuperscript{\hyperlink{a}{*}} & 41.12 $\pm$ 2.41\textsuperscript{\hyperlink{a}{*}} & 19.85 $\pm$ 1.38\textsuperscript{\hyperlink{a}{*}} \\
			ScribbleVC~\cite{li2023scribblevc} & 62.50 $\pm$ 1.61\textsuperscript{\hyperlink{a}{*}} & 47.93 $\pm$ 1.69\textsuperscript{\hyperlink{a}{*}} & 39.89 $\pm$ 2.78 & 19.32 $\pm$ 1.52\textsuperscript{\hyperlink{a}{*}} \\
			ScribFromer~\cite{li2024scribformer} & 59.68 $\pm$ 1.95\textsuperscript{\hyperlink{a}{*}} & 45.08 $\pm$ 1.93\textsuperscript{\hyperlink{a}{*}} & 43.15 $\pm$ 2.53\textsuperscript{\hyperlink{b}{**}} & 21.50 $\pm$ 1.41\textsuperscript{\hyperlink{a}{*}} \\
			DMSPS~\cite{han2024dmsps} & 63.22 $\pm$ 1.29\textsuperscript{\hyperlink{a}{*}} & 48.87 $\pm$ 1.20\textsuperscript{\hyperlink{a}{*}} & 42.93 $\pm$ 3.93 & 19.70 $\pm$ 0.77\textsuperscript{\hyperlink{a}{*}} \\
			Bayesian\_WSS~\cite{zheng2024bayesian} & 63.43 $\pm$ 2.26 & 48.96 $\pm$ 2.49 & 36.87 $\pm$ 1.19 & 16.32 $\pm$ 0.72 \\
			Ours & \textbf{68.04 $\pm$ 3.58} & \textbf{54.18 $\pm$ 4.05} & \textbf{34.69 $\pm$ 3.67} & \textbf{14.79 $\pm$ 2.56} \\
			\bottomrule
			\multicolumn{5}{l}{\hypertarget{a}{\textcolor{blue}{*}} The result is significantly different from ours with $p < 0.05$ via paired t-test. \hypertarget{b}{\textcolor{blue}{**}} $p < 0.01$.}
	\end{tabular}}
\end{center}
\end{table*}

\begin{figure*}[!t]
	\includegraphics[width=1\linewidth]{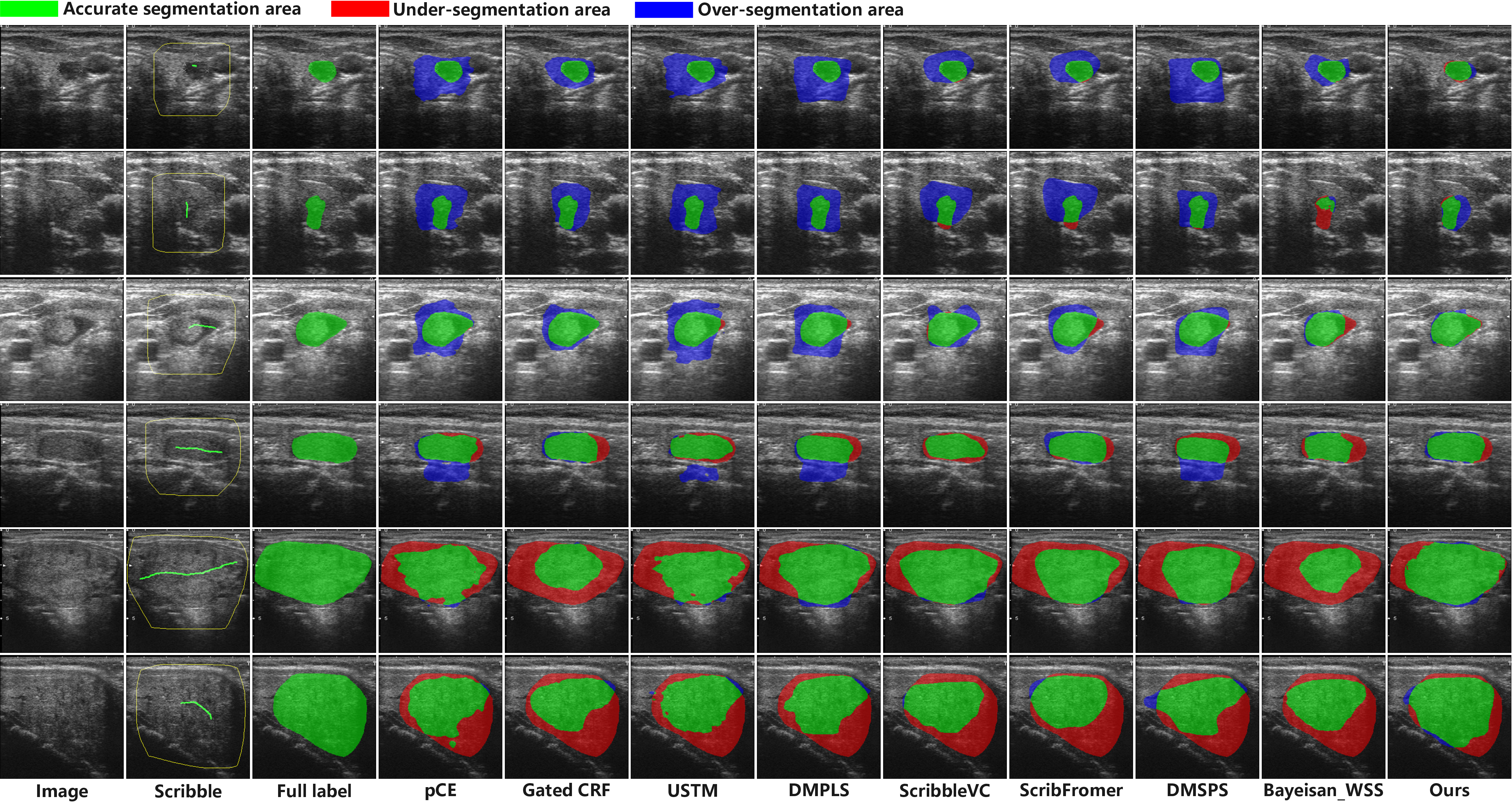}
	\caption{Visualization of comparison experimental results on the DDTI dataset. The \textit{Scribble} column represents scribble annotations, while the \textit{Full label} column indicates full dense annotations. Green arrows point out the missegmented regions. In scribble annotations, the yellow area represents unlabeled area. In full labels, the green area denotes the ground truth. In other segmentation results, green area indicates accurate segmentation, i.e., the overlap between the segmentation result and the ground truth. Red area represents under-segmentation (false negatives), while blue area indicate over-segmentation (false positives).}
	\label{fig:DDTI}
\end{figure*}

\begin{table*}[!t]
\begin{center}
	\caption{Proportion of samples where our method achieves higher Dice scores than each of the other methods across the four datasets.}
	\label{tab:num_best_results}
	\setlength{\tabcolsep}{4mm}{
		\begin{tabular}{ c | c  c  c  c }
			\toprule
			Methods & CardiacUDA & EchoNet & BUSI & DDTI \\
			\midrule
			U-Net + pCE			& 88.98 \% & 96.04 \% & 73.57 \% & 55.26 \% \\
			U-Net + Gated CRF	& 42.84 \% & 64.71 \% & 56.11 \% & 50.71 \% \\
			U-Net + USTM		& 88.49 \% & 94.81 \% & 74.81 \% & 56.99 \% \\
			DMPLS				& 88.53 \% & 91.85 \% & 66.00 \% & 58.71 \% \\
			ScribbleVC			& 68.80 \% & 50.52 \% & 65.84 \% & 62.01 \% \\
			ScribFromer			& 89.91 \% & 98.07 \% & 80.68 \% & 70.02 \% \\
			DMSPS				& 72.40 \% & 77.29 \% & 60.28 \% & 61.22 \% \\
			Bayesian\_WSS		& 49.02 \% & 88.66 \% & 61.67 \% & 63.11 \% \\
			\bottomrule
	\end{tabular}}
\end{center}
\end{table*}

This subsection compared our proposed MambaEviScrib with 5 advanced scribble-supervised learning methods on 4 diverse datasets. The methods encompass pCE (Tang \textit{et al.}, 2018)~\cite{tang2018normalized}, Gated CRF (Obukhov \textit{et al.}, 2019)~\cite{obukhov2019gated}, USTM (Liu \textit{et al.}, 2022)~\cite{liu2022weakly}, DMPLS (Luo \textit{et al.}, 2022)~\cite{luo2022scribble}, ScribbleVC (Li \textit{et al.}, 2023)~\cite{li2023scribblevc}, ScribFromer (Li \textit{et al.}, 2024)~\cite{li2024scribformer}, DMSPS (Han \textit{et al.}, 2024)~\cite{han2024dmsps}, and Bayesian\_WSS (Zheng \textit{et al.}, 2024)~\cite{zheng2024bayesian}. pCE, serving as the baseline, set the lower bound for all approaches. The backbone of pCE, Gated CRF, USTM, DMPLS, DMSPS, and Bayesian\_WSS is the U-Net architecture, while both ScribbleVC and ScribFromer incorporate Transformers. Additionally, a fully supervised U-Net~\cite{ronneberger2015u} was trained to serve as an upper reference. It is noteworthy that while our method incorporated U-Net and Mamba-UNet, only U-Net was required during testing or inference.

\subsubsection{Results on the CardiacUDA dataset}
The quantitative results of all methods obtained from the experiments on the challenging CardiacUDA dataset are presented in Table~\ref{tab:CardiacUDA}. As can be seen from the table, The baseline method, pCE, performed mediocrely and served as a benchmark for scribble-supervised segmentation performance. The performance of all other methods surpassed that of pCE. Among the existing methods, Gated CRF exhibited commendable performance, ranking second only to the proposed approach, with the mean Dice score of 74.62\% and the 95HD of 11.74. This suggested that Gated CRF holds an advantage in the task of scribble-supervised segmentation on ultrasound datasets, attributed to its capability of perceiving morphological features of anatomical structures. Our proposed method achieved the highest performance on all evaluation metrics for the scribble-supervised task on this dataset, with the mean Dice score of 75.39\% and the 95HD of 11.60, outperforming Gated CRF by 0.77\% in Dice score. This may stem from the model's tendency to prioritize learning easily segmentable classes, thereby introducing unfairness. In contrast, our method not only achieved overall superior segmentation performance but also demonstrated a narrower performance gap among different classes, highlighting its superiority and robustness.

We visualized the segmentation results as shown in Figure~\ref{fig:CardiacUDA}, aiming to provide a clear and intuitive understanding of the performance differences between various methods. As seen from the figure, the images in the first and second rows exhibited higher quality and clearer boundaries compared to other rows, resulting in overall better segmentation outcomes. Most segmentation methods committed errors in segmenting the images in the third and fourth rows, whereas our method avoided the severe mistake of misidentifying anatomical structural locations. The sixth row shows that other methods failed to segment the left atrium, whereas our approach successfully achieved it. In relative terms, pCE, USTM, and DMPLS exhibited a higher number of segmentation error regions. From the figure, we observed that the most prevalent error was the misidentification of other cardiac chambers as the left atrium, likely attributed to the small proportion of the left atrium's area in most echocardiographic images, rendering it challenging for weakly supervised models to adequately capture the characteristics of the left atrium. Furthermore, our observations revealed that ScribbleVC tended to adopt a conservative approach in segmenting certain regions, such as the left and right ventricles in the figure, failing to fully delineate the target objects. Conversely, it exhibited a more aggressive segmentation in other regions, like the left and right atria in the figure, by erroneously including substantial non-target areas within the segmentation. While it has been reported to excel in magnetic resonance image segmentation in the literature~\cite{li2023scribblevc}, it may not be adaptable to the segmentation of ultrasound images.

\begin{table*}[ht]
	\begin{center}
		\caption{Experimental results of USFM fine-tuned and tested on our four segmentation datasets.}
		\label{tab:USFM}
		\setlength{\tabcolsep}{3mm}{
			\begin{tabular}{ c | c  c  c  c }
				\toprule
				Datasets & Dice (\%) $\uparrow$ & Jaccard (\%) $\uparrow$ & 95HD (pixel) $\downarrow$ & ASD (pixel) $\downarrow$ \\
				\midrule
				CardiacUDA & 71.90 $\pm$ 1.66 & 63.26 $\pm$ 2.62 & 14.78 $\pm$ 6.09 & 6.62 $\pm$ 3.70 \\
				EchoNet & 91.19 $\pm$ 0.15 & 84.26 $\pm$ 0.28 & 3.40 $\pm$ 0.12 & 1.24 $\pm$ 0.05 \\
				BUSI & 74.20 $\pm$ 3.27 & 63.39 $\pm$ 3.18 & 33.81 $\pm$ 3.52 & 13.09 $\pm$ 1.74 \\
				DDTI & 65.12 $\pm$ 3.58 & 50.95 $\pm$ 4.05 & 36.53 $\pm$ 3.72 & 16.43 $\pm$ 2.56 \\
				\bottomrule
		\end{tabular}}
	\end{center}
\end{table*}

\begin{table*}[!t]
	\begin{center}
		\caption{Experimental results of MOFO directly tested on our three single-class segmentation datasets.}
		\label{tab:MOFO}
		\setlength{\tabcolsep}{3mm}{
			\begin{tabular}{ c | c  c  c  c }
				\toprule
				Datasets & Dice (\%) $\uparrow$ & Jaccard (\%) $\uparrow$ & 95HD (pixel) $\downarrow$ & ASD (pixel) $\downarrow$ \\
				\midrule
				EchoNet & 9.38 $\pm$ 0.15 & 5.60 $\pm$ 0.11 & 26.36 $\pm$ 0.11 & 6.22 $\pm$ 0.04 \\
				BUSI & 89.70 $\pm$ 1.09 & 83.00 $\pm$ 1.20 & 10.94 $\pm$ 2.00 & 3.67 $\pm$ 0.79 \\
				DDTI & 90.46 $\pm$ 0.76 & 83.68 $\pm$ 0.91 & 11.49 $\pm$ 0.67 & 4.19 $\pm$ 0.43 \\
				\bottomrule
		\end{tabular}}
	\end{center}
\end{table*}

In summary, our MambaEviSrib demonstrated promising performance in the multi-class segmentation of echocardiography, featuring a reduction in critical errors and generally improved segmentation accuracy. Moreover, our method depicted edge details more accurately, indicating its perception of global morphological characteristics and fusion of global and local information.

\subsubsection{Results on the EchoNet dataset}
We further conducted a comparation experiment on another echocardiography dataset EchoNet, and the quantitative results are shown in Table~\ref{tab:EchoNet}. The pCE remained as the baseline method. Surprisingly, however, ScribbleFormer performed poorly on this dataset, even falling below the lower bound. This is likely attributed to the limited annotation information, which prevented it from accurately capturing the characteristics of segmentation objects, such as morphology and edges. The majority of the remaining methods surpassed the baseline approach, with ScribbleVC achieving good performance, ranking second only to our method, with Dice scores and 95HD of 84.77\% and 6.01 respectively. In contrast to its performance on the CardiacUDA dataset, on this dataset, ScribbleVC surpassed the Gated CRF method, relegating the latter to third place. Our method demonstrated optimal performance, achieving 85.10\% Dice and 5.95 in 95HD, respectively. While there existed a gap compared to fully supervised methods, it is important to note that this was achieved under the premise of extremely sparse scribble annotations, which constituted a minimal fraction of the segmented objects, posing significant challenges. Furthermore, confronted with the challenge of varying left ventricular shapes in the EchoNet dataset, our method exhibited satisfactory performance, indicating its capability to capture and fuse global and local information, as well as morphological features.

\begin{table*}[ht]
	\begin{center}
		\caption{Ablation experiment results of the dual-branch network on the BUSI dataset. The results in bold are the best. FLOPs represent the floating point operations of the Branch-2 model, and Time indicates the duration it takes for the Branch-2 model to input and process a single image.}
		\label{tab:ablation_mamba_BUSI}
		\setlength{\tabcolsep}{2.5mm}{
			\begin{tabular}{ c  c | c  c  c  c  c }
				\toprule
				Branch-1 & Branch-2 & Dice (\%) $\uparrow$ & 95HD (pixel) $\downarrow$ & Params (M) $\downarrow$ & FLOPs (G) $\downarrow$ & Time (ms) $\downarrow$ \\
				\midrule
				U-Net & HRNet & 72.92 $\pm$ 3.57 & 35.23 $\pm$ 4.80 & \textbf{9.64} & 4.65 & 38.33 $\pm$ 0.33 \\
				U-Net & TransUNet & 73.22 $\pm$ 3.40 &  33.67 $\pm$ 4.29 & 105.32 & 33.41 & 16.33 $\pm$ 0.43 \\
				U-Net & Swin-UNet & 72.52 $\pm$ 3.84 & 32.91 $\pm$ 5.41 & 41.39 & 11.37 & 15.72 $\pm$ 0.51 \\
				U-Net & Mamba-UNet & \textbf{74.07 $\pm$ 3.27} & \textbf{31.80 $\pm$ 2.73} & 19.12 & \textbf{4.56} & \textbf{9.59 $\pm$ 0.14} \\
				\bottomrule
		\end{tabular}}
	\end{center}
\end{table*}

\begin{table*}[!t]
	\begin{center}
		\caption{Ablation experiment results of the dual-branch network on the EchoNet and DDTI datasets. The results in bold are the best.}
		\label{tab:ablation_mamba_EchoNet_DDTI}
		\setlength{\tabcolsep}{4mm}{
			\begin{tabular}{ c  c | c  c | c  c }
				\toprule
				\multirow{2}{*}{Branch-1} & \multirow{2}{*}{Branch-2} & Dice (\%) $\uparrow$ & 95HD (pixel) $\downarrow$ & Dice (\%) $\uparrow$ & 95HD (pixel) $\downarrow$ \\
				\cmidrule(lr){3-6}
				& & \multicolumn{2}{c|}{EchoNet} & \multicolumn{2}{c}{DDTI} \\
				\midrule
				U-Net & HRNet & 83.52 $\pm$ 0.68 & 6.42 $\pm$ 0.31 & 65.23 $\pm$ 3.45 & 36.85 $\pm$ 3.12 \\
				U-Net & TransUNet & 84.13 $\pm$ 0.55 & 6.25 $\pm$ 0.28 & 66.74 $\pm$ 3.21 & 35.91 $\pm$ 3.58 \\
				U-Net & Swin-UNet & 83.89 $\pm$ 0.62 & 6.35 $\pm$ 0.34 & 66.12 $\pm$ 3.67 & 35.14 $\pm$ 3.24 \\
				U-Net & Mamba-UNet & \textbf{85.10 $\pm$ 0.52} & \textbf{5.95 $\pm$ 0.24} & \textbf{68.04 $\pm$ 3.58} & \textbf{34.69 $\pm$ 3.67} \\
				\bottomrule
		\end{tabular}}
	\end{center}
\end{table*}

\begin{table*}[!t]
	\begin{center}
		\caption{Ablation experiment results of the dual-branch network on the CardiacUDA dataset. The results in bold are the best.}
		\label{tab:ablation_mamba_CardiacUDA}
		\setlength{\tabcolsep}{1.4mm}{
			\begin{tabular}{ c | c  c | c  c  c  c  c }
				\toprule
				Metrics & Branch-1 & Branch-2 & Left ventricle & Left atrium & Right atrium & Right ventricle & Mean \\
				\midrule
				\multirow{4}*{\makecell[c]{Dice \\ (\%) $\uparrow$}} & U-Net & HRNet & 73.68 $\pm$ 1.42 & 74.10 $\pm$ 0.90 & 76.94 $\pm$ 0.54 & 73.95 $\pm$ 0.53 & 74.67 $\pm$ 1.59 \\
				& U-Net & TransUNet & 74.03 $\pm$ 0.81 & 73.72 $\pm$ 0.42 & 76.77 $\pm$ 0.60 & 73.83 $\pm$ 0.48 & 74.59 $\pm$ 1.41 \\
				& U-Net & Swin-UNet & 73.97 $\pm$ 0.14 & 73.73 $\pm$ 0.74 & 77.27 $\pm$ 0.47 & 73.31 $\pm$ 0.51 & 74.57 $\pm$ 1.70 \\
				& U-Net & Mamba-UNet & \textbf{74.36 $\pm$ 0.44} & \textbf{75.16 $\pm$ 1.07} & \textbf{77.85 $\pm$ 0.77} & \textbf{74.19 $\pm$ 0.79} & \textbf{75.39 $\pm$ 1.67} \\
				\midrule
				\multirow{4}*{\makecell[c]{95HD \\ (pixel) $\downarrow$}} & U-Net & HRNet & 18.67 $\pm$ 0.13 & 10.90 $\pm$ 0.95 & 7.63 $\pm$ 0.54 & \textbf{9.48 $\pm$ 0.33} & 11.67 $\pm$ 4.42 \\
				& U-Net & TransUNet & 18.63 $\pm$ 0.13 & 11.07 $\pm$ 1.37 & 7.89 $\pm$ 0.50 & 9.71 $\pm$ 0.29 & 11.83 $\pm$ 4.31 \\
				& U-Net & Swin-UNet & 19.50 $\pm$ 0.98 & 10.72 $\pm$ 0.81 & \textbf{7.56 $\pm$ 0.44} & 10.32 $\pm$ 0.72 & 12.03 $\pm$ 4.73 \\
				& U-Net & Mamba-UNet & \textbf{18.57 $\pm$ 0.39} & \textbf{10.56 $\pm$ 1.41} & 7.64 $\pm$ 0.37 & 9.64 $\pm$ 0.12 & \textbf{11.60 $\pm$ 4.33} \\
				\bottomrule
		\end{tabular}}
	\end{center}
\end{table*}

All segmentation results are visualized in Figure~\ref{fig:EchoNet}. Six sets of echocardiographic images of the left ventricle with varying sizes were selected, and arranged from top to bottom in ascending order of scale. The green arrows in the figure pointed out the regions where segmentation errors were apparent. As can be seen from the figure, some methods tended to be aggressive in segmentation, especially when dealing with small left ventricles. In the case of large-sized left ventricles, methods like USTM and DMPLS displayed a pronounced tendency towards conservative segmentation, significantly compromising the precision of the anatomical delineation achieved. This suggests that these methods had an insufficient capture of global information. Among the two methods, ScribFormer's segmented regions notably exhibited a tendency towards aggressiveness in comparison to DMPLS, which we hypothesize may stem from the influence of the Transformer architecture. Nonetheless, they remained insufficient in adequately capturing the morphological characteristics. Furthermore, we notice that some methods failed to segment the region covered by the mitral valve, excluding it from the left ventricle. We speculate that this could be due to the insufficient coverage of scribble annotations in this region, resulting in the model's inability to capture its features. In summary, our method can overcome most of the above problems and demonstrate advantages in ultrasonic data.

\subsubsection{Results on the BUSI dataset}
In addition to echocardiography data, we also conducted comparison experiments on the breast ultrasound dataset BUSI, with the results presented in Table~\ref{tab:BUSI}. Among the existing approaches, the DMSPS method exhibited the best performance, achieving a Dice score of 71.75\%, significantly outperforming the baseline method pCE. Our proposed method, however, surpassed the DMSPS by a notable margin, attaining a Dice score of 74.07\%, which represented a 2.32\% improvement over DMSPS and further narrowed the gap towards the fully supervised upper bound of 77.91\%. Our method exhibited the most significant improvement on this dataset among the four, indicating a certain advantage in breast lesion detection.

\begin{table*}[ht]
	\begin{center}
		\caption{Ablation experiment results of the consistency strategy on the CardiacUDA dataset. EDL stands for evidential deep learning, while EGC represents evidence-guided consistency strategy. The results in bold are the best.}
		\label{tab:ablation_consistency_CardiacUDA}
		\setlength{\tabcolsep}{2.6mm}{
			\begin{tabular}{ c | c  c | c  c  c  c  c }
				\toprule
				Metrics & EDL & EGC & Left ventricle & Left atrium & Right atrium & Right ventricle & Mean \\
				\midrule
				\multirow{3}*{\makecell[c]{Dice \\ (\%) $\uparrow$}} & & & 74.31 $\pm$ 0.23 & 73.69 $\pm$ 1.47 & 76.44 $\pm$ 0.83 & 71.16 $\pm$ 0.59 & 73.90 $\pm$ 2.11 \\
				& \Checkmark & & 74.05 $\pm$ 0.31 & 73.46 $\pm$ 1.24 & 77.10 $\pm$ 0.40 & 74.06 $\pm$ 0.83 & 74.67 $\pm$ 1.63 \\
				& \Checkmark & \Checkmark & \textbf{74.36 $\pm$ 0.44} & \textbf{75.16 $\pm$ 1.07} & \textbf{77.85 $\pm$ 0.77} & \textbf{74.19 $\pm$ 0.79} & \textbf{75.39 $\pm$ 1.67} \\
				\midrule
				\multirow{3}*{\makecell[c]{95HD \\ (pixel) $\downarrow$}} & & & \textbf{18.39 $\pm$ 0.26} & 10.85 $\pm$ 1.25 & 8.48 $\pm$ 1.04 & 15.94 $\pm$ 0.15 & 13.42 $\pm$ 4.17 \\
				& \Checkmark & & 18.82 $\pm$ 0.30 & 10.86 $\pm$ 0.67 & 7.86 $\pm$ 0.32 & \textbf{9.56 $\pm$ 0.35} & 11.77 $\pm$ 4.41 \\
				& \Checkmark & \Checkmark & 18.57 $\pm$ 0.39 & \textbf{10.56 $\pm$ 1.41} & \textbf{7.64 $\pm$ 0.37} & 9.64 $\pm$ 0.12 & \textbf{11.60 $\pm$ 4.33} \\
				\bottomrule
		\end{tabular}}
	\end{center}
\end{table*}

\begin{table*}[!t]
	\begin{center}
		\caption{Ablation experiment results of the consistency strategy on the EchoNet, BUSI, and DDTI datasets. EDL stands for evidential deep learning, while EGC represents evidence-guided consistency strategy. The results in bold are the best.}
		\label{tab:ablation_consistency}
		\setlength{\tabcolsep}{2mm}{
			\begin{tabular}{ c  c | c  c | c  c | c  c }
				\toprule
				\multirow{2}{*}{EDL} & \multirow{2}{*}{EGC} & Dice (\%) $\uparrow$ & 95HD (pixel) $\downarrow$ & Dice (\%) $\uparrow$ & 95HD (pixel) $\downarrow$ & Dice (\%) $\uparrow$ & 95HD (pixel) $\downarrow$ \\
				\cmidrule(lr){3-8}
				& & \multicolumn{2}{c|}{EchoNet} & \multicolumn{2}{c|}{BUSI} & \multicolumn{2}{c}{DDTI} \\
				\midrule
				& & 83.36 $\pm$ 0.65 & 6.42 $\pm$ 0.30 & 72.72 $\pm$ 3.35 & 32.73 $\pm$ 5.02 & 66.15 $\pm$ 3.24 & 37.85 $\pm$ 3.56 \\
				\Checkmark & & 84.15 $\pm$ 0.58 & 6.22 $\pm$ 0.35 & 73.59 $\pm$ 2.76 & 33.41 $\pm$ 7.21 & 66.82 $\pm$ 3.10 & 39.91 $\pm$ 3.72 \\
				\Checkmark & \Checkmark & \textbf{85.10 $\pm$ 0.52} & \textbf{5.95 $\pm$ 0.24} & \textbf{74.07 $\pm$ 3.27} & \textbf{31.10 $\pm$ 2.73} & \textbf{68.04 $\pm$ 3.58} & \textbf{34.69 $\pm$ 3.67} \\
				\bottomrule
		\end{tabular}}
	\end{center}
\end{table*}

All segmentation results are visualized in Figure~\ref{fig:BUSI}. We selected a representative set of images and their corresponding segmentation outcomes, encompassing lesions of varying morphologies and sizes, for presentation. The lesions in the first row of images were relatively small, posing challenges for some methods that tended to falsely segment similar regions as lesions, leading to false positives, while others may fail to fully segment the lesions, resulting in false negatives. In the second row, the lesion is small and blends with the background in brightness, making it difficult to discern. Most methods deviated in segmentation, while our approach accurately identified the lesion's location. Furthermore, some lesions are large in size, posing challenges for most methods to achieve complete segmentation. Our approach succeeds by leveraging the model's ability to comprehend global information. Additionally, the visualization results reveal that the pCE and USTM methods produce relatively more false-negative lesions in some images of this dataset, increasing the risk of missed diagnoses.

\subsubsection{Results on the DDTI dataset}
We also conducted experiments on another challenging dataset, DDTI, for thyroid nodule ultrasound image segmentation, with the results presented in Table~\ref{tab:DDTI}. On this dataset, the Gated CRF method continued to maintain the best performance among the existing methods. Unfortunately, due to significant challenges, most methods fail to surpass even the lower-bound performance of pCE. Our method continued to outperform others, achieving a Dice score of 68.04\%, which is a 1.51\% improvement over the Gated CRF method. However, it was regrettable that there was still a certain gap from the upper bound, which was attributed to the significant challenges posed by this dataset for weak supervision.

\begin{table*}[!t]
	\begin{center}
		\caption{The ablation study results of replacing the EGC strategy with non-deep learning consistency enhancement techniques. CWGG denotes the Confidence-Weighted Gravitational Grouping algorithm, and MRF represents Markov Random Field regularization.}
		\label{tab:ablation_EGC}
		\setlength{\tabcolsep}{4mm}{
			\begin{tabular}{ c | c  c | c  c }
				\toprule
				\multirow{2}{*}{Method} & Dice (\%) $\uparrow$ & 95HD (pixel) $\downarrow$ & Dice (\%) $\uparrow$ & 95HD (pixel) $\downarrow$ \\
				\cmidrule(lr){2-5}
				& \multicolumn{2}{c|}{CardiacUDA} & \multicolumn{2}{c}{EchoNet} \\
				\midrule
				CWGG & 74.51 $\pm$ 1.54 & 12.13 $\pm$ 4.06 & 84.38 $\pm$ 0.41 & 6.11 $\pm$ 0.38 \\
				MRF & 74.02 $\pm$ 1.61 & 12.33 $\pm$ 4.24 & 83.94 $\pm$ 0.45 & 6.29 $\pm$ 0.40 \\
				All & \textbf{75.39 $\pm$ 1.67} & \textbf{11.60 $\pm$ 4.33} & \textbf{85.10 $\pm$ 0.52} & \textbf{5.95 $\pm$ 0.24} \\
				\midrule
				& \multicolumn{2}{c|}{BUSI} & \multicolumn{2}{c}{DDTI} \\
				\midrule
				CWGG & 72.48 $\pm$ 3.18 & 33.52 $\pm$ 3.20 & 65.22 $\pm$ 1.67 & 35.90 $\pm$ 2.22 \\
				MRF & 72.05 $\pm$ 3.19 & 34.13 $\pm$ 3.51 & 64.84 $\pm$ 1.76 & 36.53 $\pm$ 3.21 \\
				All & \textbf{74.07 $\pm$ 3.27} & \textbf{31.10 $\pm$ 2.73} & \textbf{68.04 $\pm$ 3.58} & \textbf{34.69 $\pm$ 3.67} \\
				\bottomrule
		\end{tabular}}
	\end{center}
\end{table*}

All segmentation results are visualized in Figure~\ref{fig:DDTI}, with the selected lesions arranged from small to large in size. Some methods tended to produce false positives for small-sized lesions and false negatives for large-sized lesions. Overall, our method exhibited better performance compared to existing methods. However, upon observing the ultrasound images from the four visualization examples, it becomes evident that the lesion boundaries are unclear, making it difficult to distinguish between the lesions and normal tissues. This poses a significant challenge for weakly supervised segmentation and underscores the need for continued focus and improvement in this area of ultrasound image segmentation in the future.

\subsubsection{Sample-wise performance dominance across four datasets}
Table~\ref{tab:num_best_results} presents the comparative superiority of our method across four ultrasound datasets, quantified as the percentage of samples where our approach achieves higher Dice scores than existing methods. Notably, our method outperforms U-Net+pCE in 88.98\%-96.04\% of CardiacUDA and EchoNet cases, demonstrating robust performance in cardiac segmentation tasks. While maintaining competitive advantages in BUSI (65.84\%-80.68\%) and DDTI (55.26\%-70.02\%), the results highlight our method's particular effectiveness in echocardiographic analysis, surpassing even transformer-based ScribFormer in 98.07\% of EchoNet samples.

\begin{figure*}[!t]
	\includegraphics[width=1\linewidth]{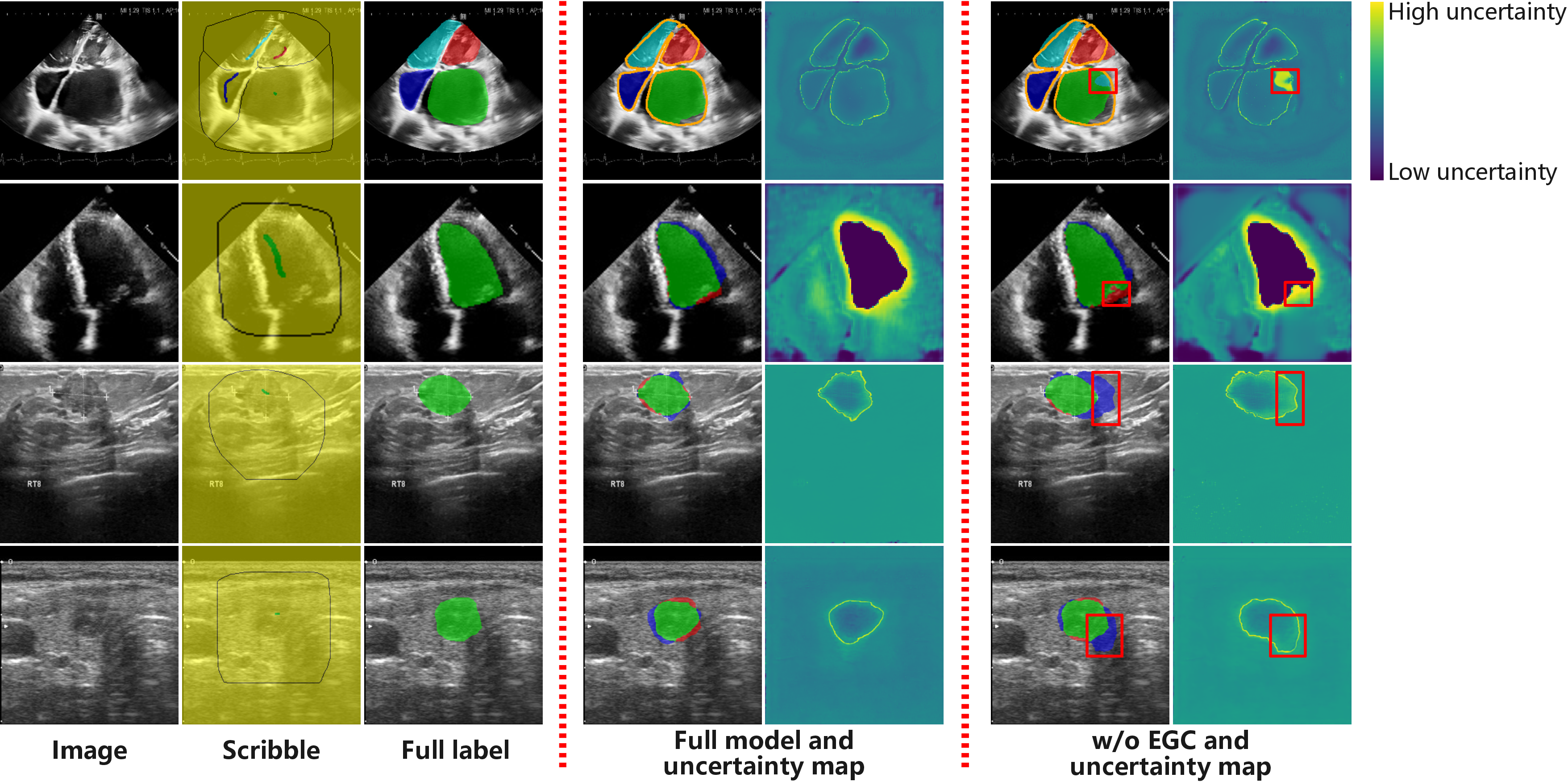}
	\caption{Visualization of ablation results for the EGC strategy, comparing segmentation outputs and uncertainty maps with/without EGC across four datasets. w/o denotes without.}
	\label{fig:ablation_consistency}
\end{figure*}

\subsubsection{Comparison with ultrasound foundation models}
We evaluated our dataset on two leading ultrasound foundation models: USFM (Jiao \textit{et al.}, 2024)~\cite{jiao2024usfm} and MOFO (Chen \textit{et al.}, 2024)~\cite{chen2024multi}. For USFM, a masked image modeling (MIM) pre-trained foundation model, we performed full fine-tuning using dense labels on our datasets. Surprisingly, as shown in Table~\ref{tab:USFM}, USFM underperformed both our weakly supervised method and the standard U-Net across all datasets, despite utilizing full supervision. EchoNet outperformed other datasets, likely due to its largest sample size, indicating that USFM's downstream fine-tuning still requires substantial data. This suggests that while foundation models demonstrate impressive generalizability, their pretraining objective of MIM may not align effectively with downstream segmentation tasks without extensive fine-tuning data. For MOFO, an ultrasound foundation model designed for single-class segmentation per inference despite its multi-organ capability, we conducted evaluations on three datasets compatible with its output structure: EchoNet (left ventricle), BUSI (breast lesions), and DDTI (thyroid nodules). The four-class CardiacUDA dataset was excluded due to structural incompatibility with MOFO's single-class output design. Using the official pre-trained MOFO model without fine-tuning, we observed limitations in its clinical applicability. Since BUSI and DDTI are included in MOFO's official training set, creating sample overlap with our test set, our test results surpass the officially reported performance, as shown in Table~\ref{tab:MOFO}. Conversely, its performance on EchoNet, containing cardiac chamber segmentation outside MOFO's training scope, highlights critical dependencies on task-specific training data coverage. These results highlight fundamental limitations of current ultrasound foundation models: 1) Performance heavily relies on domain/task alignment between pretraining and target datasets; 2) Significant performance degradation occurs when confronting unseen anatomical structures; 3) Architectural constraints limit applicability to multi-class segmentation scenarios.

\subsection{Ablation study}
\subsubsection{Comparison of the dual-branch network}
To demonstrate the superiority of Mamba-UNet within the Mamba branch of our proposed method, we conducted experiments by replacing it with other advanced networks. These networks include the Swin-UNet and TransUNet, both of which leverage the Transformer architecture, as well as the purely CNN-based HRNet. We compared their performance and efficiency on datasets. For fairness, only U-Net was utilized for inference during the testing phase. The experimental results, presented in Tables~\ref{tab:ablation_mamba_BUSI},~\ref{tab:ablation_mamba_EchoNet_DDTI} and~\ref{tab:ablation_mamba_CardiacUDA}, demonstrated that our method achieved the optimal performance with the second-least number of parameters, trailing only HRNet. However, in terms of floating-point operations (FLOPs), our method surpassed even the CNN-based HRNet, recording the lowest FLOPs at 4.56G. Furthermore, our method operated at the fastest speed, taking approximately 9.59ms to process and segment an input image of $256\times256$ pixels, which is 1.64 times faster than the second-fastest, Swin-UNet. This underscored the capability of the Mamba model to significantly reduce computational complexity while effectively capturing long-range dependencies.

\begin{table*}[ht]
	\begin{center}
		\caption{Ablation study of hyperparameters ($\tau$, $\epsilon$, and $\gamma$) across four ultrasound datasets. Bold values indicate optimal performance.}
		\label{tab:hyper_parameter}
		\setlength{\tabcolsep}{0.4mm}{
			\begin{tabular}{ c | c | c  c | c | c  c | c | c  c }
				\toprule
				Dataset & $\tau$ & Dice (\%) $\uparrow$ & 95HD (pixel) $\downarrow$ & $\epsilon$ & Dice (\%) $\uparrow$ & 95HD (pixel) $\downarrow$ & $\gamma$ & Dice (\%) $\uparrow$ & 95HD (pixel) $\downarrow$ \\
				\midrule
				\multirow{4}{*}{CardiacUDA} & 0.1 & 75.02 $\pm$ 1.84 & 11.73 $\pm$ 4.12 & 0.75 & 75.12 $\pm$ 1.84 & 11.64 $\pm$ 4.84 & 0.05 & 74.98 $\pm$ 1.65 & 11.78 $\pm$ 4.77 \\
				& 0.25 & \textbf{75.39 $\pm$ 1.67} & \textbf{11.60 $\pm$ 4.33} & 0.5 & \textbf{75.39 $\pm$ 1.67} & \textbf{11.60 $\pm$ 4.33} & 0.1 & \textbf{75.39 $\pm$ 1.67} & 11.60 $\pm$ 4.33 \\
				& 0.5 & 74.98 $\pm$ 1.72 & 11.76 $\pm$ 4.55 & 0.25 & 75.06 $\pm$ 1.58 & 11.68 $\pm$ 4.36 & 0.5 & 75.10 $\pm$ 1.83 & \textbf{11.58 $\pm$ 4.38} \\
				& 1.0 & 74.75 $\pm$ 1.59 & 11.89 $\pm$ 4.18 & 1.0 & 74.95 $\pm$ 1.73 & 11.74 $\pm$ 4.52 & 1.0 & 74.02 $\pm$ 1.80 & 11.64 $\pm$ 4.45 \\
				\midrule
				\multirow{4}{*}{EchoNet} & 0.1 & 85.04 $\pm$ 0.49 & \textbf{5.93 $\pm$ 0.25} & 0.75 & 84.98 $\pm$ 0.47 & 5.98 $\pm$ 0.25 & 0.05 & 84.93 $\pm$ 0.61 & 6.05 $\pm$ 0.32 \\
				& 0.25 & \textbf{85.10 $\pm$ 0.52} & 5.95 $\pm$ 0.24 & 0.5 & \textbf{85.10 $\pm$ 0.52} & \textbf{5.95 $\pm$ 0.24} & 0.1 & \textbf{85.10 $\pm$ 0.52} & \textbf{5.95 $\pm$ 0.24} \\
				& 0.5 & 84.96 $\pm$ 0.54 & 5.98 $\pm$ 0.32 & 0.25 & 84.94 $\pm$ 0.58 & 6.05 $\pm$ 0.27 & 0.5 & 84.89 $\pm$ 0.57 & 6.04 $\pm$ 0.28 \\
				& 1.0 & 84.88 $\pm$ 0.47 & 6.03 $\pm$ 0.30 & 1.0 & 84.82 $\pm$ 0.51 & 6.11 $\pm$ 0.29 & 1.0 & 84.77 $\pm$ 0.64 & 6.10 $\pm$ 0.34 \\
				\midrule
				\multirow{4}{*}{BUSI} & 0.1 & 73.85 $\pm$ 3.08 & 31.74 $\pm$ 2.51 & 0.75 & 73.91 $\pm$ 2.94 & 31.87 $\pm$ 2.65 & 0.05 & 73.88 $\pm$ 3.68 & 31.69 $\pm$ 2.88 \\
				& 0.25 & \textbf{74.07 $\pm$ 3.27} & \textbf{31.10 $\pm$ 2.73} & 0.5 & \textbf{74.07 $\pm$ 3.27} & 31.10 $\pm$ 2.73 & 0.1 & \textbf{74.07 $\pm$ 3.27} & \textbf{31.10 $\pm$ 2.73} \\
				& 0.5 & 73.41 $\pm$ 3.15 & 32.24 $\pm$ 2.76 & 0.25 & 73.95 $\pm$ 3.68 & \textbf{31.06 $\pm$ 2.56} & 0.5 & 73.66 $\pm$ 3.85 & 31.98 $\pm$ 2.54 \\
				& 1.0 & 72.96 $\pm$ 3.29 & 33.16 $\pm$ 2.44 & 1.0 & 73.79 $\pm$ 3.52 & 31.81 $\pm$ 2.48 & 1.0 & 73.74 $\pm$ 3.16 & 31.92 $\pm$ 2.89 \\
				\midrule
				\multirow{4}{*}{DDTI} & 0.1 & 67.78 $\pm$ 3.35 & 34.74 $\pm$ 3.15 & 0.25 & 67.65 $\pm$ 3.84 & 34.65 $\pm$ 2.98 & 0.05 & 67.81 $\pm$ 3.86 & 34.75 $\pm$ 3.84 \\
				& 0.25 & \textbf{68.04 $\pm$ 3.58} & 34.69 $\pm$ 3.67 & 0.5 & \textbf{68.04 $\pm$ 3.58} & \textbf{34.69 $\pm$ 3.67} & 0.1 & \textbf{68.04 $\pm$ 3.58} & \textbf{34.69 $\pm$ 3.67} \\
				& 0.5 & 67.69 $\pm$ 3.87 & \textbf{34.61 $\pm$ 3.40} & 0.25 & 67.51 $\pm$ 3.24 & 34.83 $\pm$ 3.65 & 0.5 & 67.70 $\pm$ 3.24 & 34.79 $\pm$ 3.71 \\
				& 1.0 & 66.98 $\pm$ 3.62 & 35.54 $\pm$ 3.46 & 1.0 & 66.73 $\pm$ 3.56 & 35.13 $\pm$ 3.58 & 1.0 & 67.02 $\pm$ 3.28 & 34.92 $\pm$ 3.51 \\
				\bottomrule
		\end{tabular}}
	\end{center}
\end{table*}

\subsubsection{Comparison of the consistency strategy}
To validate the effectiveness of EDL and EGC, ablation experiments were conducted, with results presented in Tables~\ref{tab:ablation_consistency_CardiacUDA} and~\ref{tab:ablation_consistency}. Given that EGC relies on EDL, we initially removed EGC from the framework and replaced it with a conventional mean squared error (MSE) constraint to validate the superiority of EGC. According to the experimental results, our complete method outperformed the conditions where either EGC or both EGC and EDL were removed. For instance, On the CardiacUDA dataset, the full framework achieved a mean Dice of 75.39\%, significantly outperforming configurations without EGC (74.67\%) or EDL (73.90\%), with a improvemet of 1.49\% in Dice score. Similar trends were observed in other datasets. Thereby, the proposed EGC exhibited certain advantages, facilitating the model's ability to better handle pixels located at decision boundaries, such as blurred edges in ultrasound images, thereby enhancing performance.

As shown in Figure~\ref{fig:ablation_consistency}, the ablation experiment's visual results validate the effectiveness of the EGC strategy. On the BUSI and DDTI datasets, EGC improved edge segmentation accuracy for samples with blurred lesion boundaries, with uncertainty maps indicating higher uncertainty in missegmented regions without EGC. On the EchoNet dataset, EGC enhanced mitral valve segmentation and reduced uncertainty. Additionally, the CardiacUDA experiments demonstrated that the method successfully corrected missegmentations where the left ventricle was erroneously identified as the right atrium.

Furthermore, we conducted comparative experiments replacing the EGC strategy with classical non-deep learning regularization methods to evaluate boundary uncertainty modeling. The experimental results are shown in Table~\ref{tab:ablation_EGC}. Traditional approaches, including confidence-weighted gravitational grouping (CWGG) and Markov random field (MRF) regularization, demonstrated lower performance across all datasets compared to our EGC-enhanced framework, particularly in addressing edge ambiguities inherent to ultrasound imaging. The results confirm that conventional methods lack the capability to effectively leverage predictions near decision boundaries for uncertainty refinement.

\begin{table*}[ht]
	\begin{center}
		\caption{Ablation study of loss function components across four datasets. Bold values indicate optimal performance. w/o denotes without, and L2 $\rightarrow$ KL signifies replace L2 with KL divergence.}
		\label{tab:ablation_loss}
		\setlength{\tabcolsep}{4mm}{
			\begin{tabular}{ c | c  c | c  c }
				\toprule
				\multirow{2}{*}{Configuration} & Dice (\%) $\uparrow$ & 95HD (pixel) $\downarrow$ & Dice (\%) $\uparrow$ & 95HD (pixel) $\downarrow$ \\
				\cmidrule(lr){2-5}
				& \multicolumn{2}{c|}{CardiacUDA} & \multicolumn{2}{c}{EchoNet} \\
				\midrule
				All & \textbf{75.39 $\pm$ 1.67} & \textbf{11.60 $\pm$ 4.33} & \textbf{85.10 $\pm$ 0.52} & \textbf{5.95 $\pm$ 0.24} \\
				w/o $\mathcal{L}_{ic}$ & 73.50 $\pm$ 1.89 & 13.25 $\pm$ 4.50 & 83.20 $\pm$ 0.65 & 6.42 $\pm$ 0.30 \\
				w/o $\ell_{pce}$ & 71.82 $\pm$ 2.15 & 15.30 $\pm$ 5.14 & 80.97 $\pm$ 0.72 & 7.85 $\pm$ 0.45 \\
				w/o $\mathcal{L}_{evi}$ & 74.10 $\pm$ 1.73 & 12.85 $\pm$ 4.26 & 84.05 $\pm$ 0.58 & 6.20 $\pm$ 0.35 \\
				w/o $\ell_{crf}$ & 74.85 $\pm$ 1.55 & 11.95 $\pm$ 4.15 & 84.55 $\pm$ 0.50 & 6.05 $\pm$ 0.28 \\
				L2 $\rightarrow$ KL & 74.72 $\pm$ 1.62 & 12.10 $\pm$ 4.32 & 84.80 $\pm$ 0.48 & 6.00 $\pm$ 0.25 \\
				\midrule
				& \multicolumn{2}{c|}{BUSI} & \multicolumn{2}{c}{DDTI} \\
				\midrule
				All & \textbf{74.07 $\pm$ 3.27} & \textbf{31.10 $\pm$ 2.73} & \textbf{68.04 $\pm$ 3.58} & \textbf{34.69 $\pm$ 3.67} \\
				w/o $\mathcal{L}_{ic}$ & 72.50 $\pm$ 3.45 & 33.25 $\pm$ 2.98 & 66.15 $\pm$ 3.24 & 37.85 $\pm$ 3.50 \\
				w/o $\ell_{pce}$ & 70.83 $\pm$ 3.12 & 35.40 $\pm$ 3.15 & 63.93 $\pm$ 3.45 & 40.10 $\pm$ 3.85 \\
				w/o $\mathcal{L}_{evi}$ & 72.92 $\pm$ 3.23 & 32.78 $\pm$ 2.67 & 66.80 $\pm$ 3.10 & 39.91 $\pm$ 3.72 \\
				w/o $\ell_{crf}$ & 73.26 $\pm$ 3.08 & 31.51 $\pm$ 2.89 & 67.25 $\pm$ 3.05 & 35.20 $\pm$ 3.55 \\
				L2 $\rightarrow$ KL & 73.57 $\pm$ 2.95 & 31.20 $\pm$ 2.75 & 67.54 $\pm$ 3.15 & 34.95 $\pm$ 3.60 \\
				\bottomrule
		\end{tabular}}
	\end{center}
\end{table*}

\subsubsection{Hyper-parameter analysis}
We conducted an analysis on 3 hyper-parameters that require manual tuning, namely $\tau$, $\epsilon$, and $\gamma$. The parameter $\tau$ served as a scaling factor in the transformation function $f_e(\cdot)$ (see Eq.~\eqref{eq:transform_function}), modulating the smoothness of the function. $\epsilon$, as the sharpening temperature in Eqs.~\eqref{eq:sharping_temperature_cnn} and~\eqref{eq:sharping_temperature_mamba}, biases predictions of high evidence towards regions of higher density. The parameter $\gamma$ in Eqs.~\eqref{eq:loss_weight_cnn} and~\eqref{eq:loss_weight_mamba} served as a balancing weight between the loss functions pCE and Gated CRF. The experimental results in Table~\ref{tab:hyper_parameter} indicate that the model achieves optimal performance when $\tau$, $\epsilon$, and $\gamma$ are set to 0.25, 0.5, and 0.1, respectively. Notably, while these parameters demonstrate dataset-agnostic stability (optimal values remain consistent across echocardiography and lesion segmentation tasks), slight performance variations suggest that task-specific tuning could further enhance results. For instance, BUSI benefits marginally from $\tau=0.5$ due to its higher lesion-background contrast, while DDTI shows improved boundary precision with $\epsilon=0.25$ to accommodate its ambiguous edges. This analysis confirms the robustness of our parameter configuration.

\subsubsection{Ablation study on loss function components}
To validate the contributions of different loss components, we conducted comprehensive ablation studies across four datasets, as shown in Table~\ref{tab:ablation_loss}. Removing the evidence-guided consistency loss $\mathcal{L}_{ic}$ caused significant performance degradation, reducing Dice scores by 1.89\% (CardiacUDA), 1.90\% (EchoNet), 1.57\% (BUSI), and 1.89\% (DDTI), highlighting its critical role in refining boundary predictions. Excluding the partial cross-entropy loss $\ell_{pce}$ led to even more severe declines (e.g., 3.57\% Dice drop on CardiacUDA), underscoring its necessity for leveraging sparse scribble supervision. The evidential loss $\mathcal{L}_{evi}$ removal moderately impacted performance (1.29\%-2.15\% Dice reduction), confirming its effectiveness in uncertainty modeling. While omitting the Gated CRF loss $\ell_{crf}$ showed minimal impact, replacing the L2 loss in EGC with KL divergence slightly reduced performance, validating the superiority of the L2-based guidance strategy. These experiments collectively demonstrate the synergistic importance of all proposed loss components, with $\mathcal{L}_{ic}$ and $\ell_{pce}$ being particularly vital for robust weakly supervised learning.

\begin{table*}[ht]
	\begin{center}
		\caption{Generalization evaluation of models. The models trained on the CardiacUDA dataset are tested on the EchoNet dataset.}
		\label{tab:generalization}
		\setlength{\tabcolsep}{4mm}{
			\begin{tabular}{ c | c  c  c  c }
				\toprule
				Methods & Dice (\%) $\uparrow$ & Jaccard (\%) $\uparrow$ & 95HD (pixel) $\downarrow$ & ASD (pixel) $\downarrow$ \\
				\midrule
				U-Net + pCE & 66.76 $\pm$ 6.65 & 53.57 $\pm$ 6.59 & 13.01 $\pm$ 1.95 & 4.31 $\pm$ 0.54 \\
				U-Net + Gated CRF & 74.13 $\pm$ 0.61 & \textit{61.87 $\pm$ 0.49} & \textit{11.14 $\pm$ 0.34} & \textit{3.26 $\pm$ 0.21} \\
				U-Net + USTM & 63.08 $\pm$ 7.69 & 49.89 $\pm$ 7.60 & 13.95 $\pm$ 1.25 & 4.57 $\pm$ 0.18 \\
				DMPLS & 60.07 $\pm$ 3.31 & 46.93 $\pm$ 2.88 & 15.33 $\pm$ 0.84 & 4.59 $\pm$ 0.17 \\
				ScribbleVC & 64.87 $\pm$ 1.95 & 51.99 $\pm$ 1.96 & 13.80 $\pm$ 0.66 & 3.75 $\pm$ 0.30 \\
				ScribFromer & 66.35 $\pm$ 4.09 & 52.69 $\pm$ 4.49 & 15.28 $\pm$ 1.34 & 5.36 $\pm$ 0.77 \\
				DMSPS & \textit{74.49 $\pm$ 2.89} & 61.47 $\pm$ 2.26 & 11.22 $\pm$ 1.17 & 3.53 $\pm$ 0.44 \\
				Bayesian\_WSS & 70.51 $\pm$ 4.72 & 58.64 $\pm$ 4.49 & 11.90 $\pm$ 1.53 & \textbf{3.20 $\pm$ 0.04} \\
				Ours & \textbf{76.39 $\pm$ 1.64} & \textbf{64.22 $\pm$ 1.66} & \textbf{10.82 $\pm$ 0.54} & 3.37 $\pm$ 0.10 \\
				\bottomrule
		\end{tabular}}
	\end{center}
\end{table*}

\subsubsection{Ablation study on threshold $\lambda_{iter}$}
\begin{figure*}[ht]
	\includegraphics[width=1\linewidth]{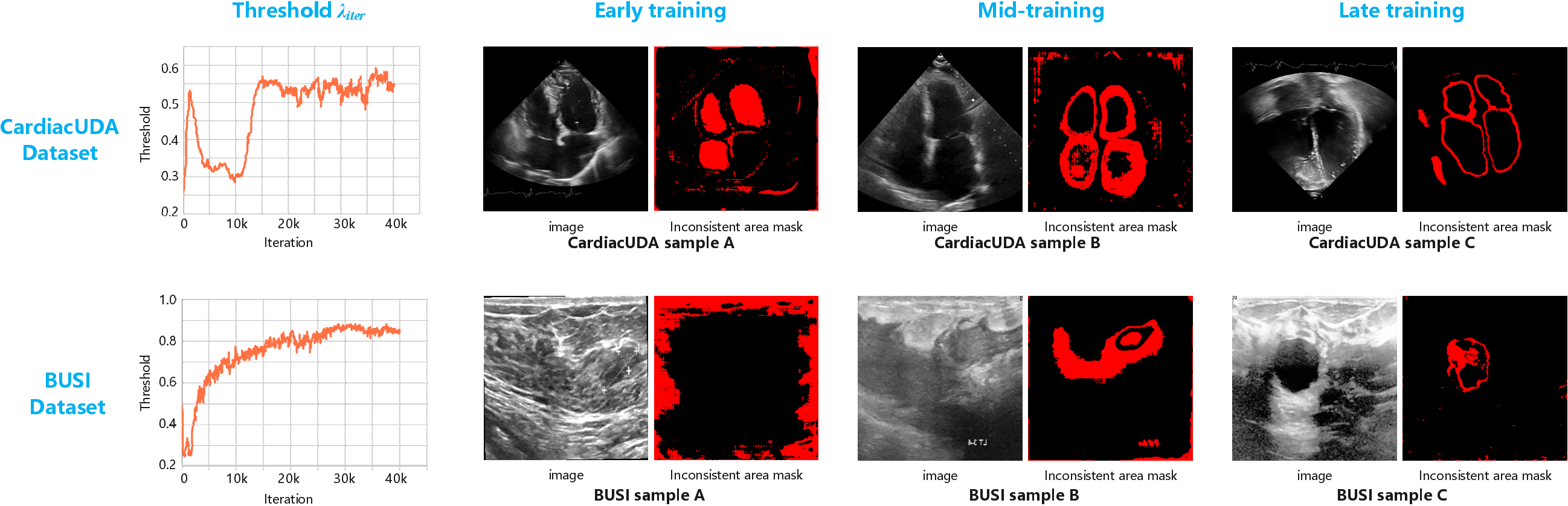}
	\caption{The left tables illustrate the variation of threshold $\lambda_{iter}$ across training iterations on the CardiacUDA and BUSI datasets. The right side displays samples from different training epochs for both datasets, along with their corresponding inconsistency region masks generated based on $\lambda_{iter}$.}
	\label{fig:ablation_thr}
\end{figure*}

$\lambda_{iter}$ is used in this paper to distinguish consistent and inconsistent regions between the outputs of the two network branches. Consistent regions are optimized using traditional cross-entropy supervision, while inconsistent regions are optimized using our EGC strategy. In the early training phase, pseudo-labels are often unstable and may contain numerous segmentation errors. A low initial threshold encourages the model to generate diverse pseudo-labels, exploring unannotated regions. If the initial threshold is too high, with too few high-confidence regions, EGC lacks sufficient "reliable anchors" to guide the optimization of low-confidence regions. Thus, retaining more uncertain regions early on prevents the model from converging to local optima by prematurely relying on a limited set of high-confidence samples. As shown in Figure~\ref{fig:ablation_thr}, the threshold initially remains low, gradually increases with training iterations, and eventually stabilizes, aligning with expectations. We present the masks of inconsistent regions generated by $\lambda_{iter}$ at different training stages. Early in training, the inconsistent masks appear irregular but progressively shrink to the edges of regions of interest as training advances, as these areas typically lie near decision boundaries.

Furthermore, we introduced a comparison between low and high thresholds. Given that varying numbers of classes or datasets lead to different distributions in uncertainty maps, the thresholds are adaptive. We set the low threshold as $m/2$ and the high threshold as $(1+m)/2$, based on the mean $m$ of the uncertainty map. Experimental results on CardiacUDA and BUSI, as shown in Table~\ref{tab:ablation_thr}, indicate that while the fluctuations are minimal, our threshold-setting approach outperforms others.

\begin{table*}[!t]
	\begin{center}
		\caption{Ablation study results comparing various $\lambda_{iter}$ threshold configurations on the CardiacUDA and BUSI datasets.}
		\label{tab:ablation_thr}
		\setlength{\tabcolsep}{6mm}{
			\begin{tabular}{ c | c  c | c  c }
				\toprule
				\multirow{2}{*}{$\lambda_{iter}$} & Dice (\%) $\uparrow$ & 95HD (pixel) $\downarrow$ & Dice (\%) $\uparrow$ & 95HD (pixel) $\downarrow$ \\
				\cmidrule(lr){2-5}
				& \multicolumn{2}{c|}{CardiacUDA} & \multicolumn{2}{c}{BUSI} \\
				\midrule
				Low & 74.89 $\pm$ 1.52 & 12.35 $\pm$ 4.21 & 73.29 $\pm$ 3.35 & 31.85 $\pm$ 2.91 \\
				High & 74.34 $\pm$ 1.88 & 11.92 $\pm$ 4.43 & 73.04 $\pm$ 3.42 & 32.15 $\pm$ 3.05 \\
				Ours & \textbf{75.39 $\pm$ 1.67} & \textbf{11.60 $\pm$ 4.33} & \textbf{74.07 $\pm$ 3.27} & \textbf{31.30 $\pm$ 2.73} \\
				\bottomrule
		\end{tabular}}
	\end{center}
\end{table*}

\subsection{Generalization and robustness analysis}
\subsubsection{Generalization analysis}
To assess the generalization capabilities of each model, we subjected the models trained on the CardiacUDA dataset to testing on the EchoNet dataset, which served as an out-of-distribution (OOD) dataset. This is crucial when applying the models in real-world clinical scenarios. As the EchoNet dataset is dedicated solely to segmenting the left ventricle, whereas the CardiacUDA dataset encompassed all four cardiac chambers including the left ventricle, only the left ventricular segmentation results were retained during testing on the EchoNet dataset. Furthermore, given that the images in the CardiacUDA dataset are of size 256$\times$256 pixels, whereas those in the EchoNet dataset are 112$\times$112 pixels, all images were first upscaled to 256$\times$256 pixels before testing on the EchoNet dataset. Subsequently, the obtained prediction masks were downscaled to 112$\times$112 pixels, upon which the evaluation metrics were calculated. The evaluation results are presented in Table~\ref{tab:generalization}. The findings indicate that among the existing methods, the Gated CRF method performed the best, while the other existing methods even underperformed the baseline pCE method, suggesting their inadequate generalization capabilities and possible overfitting issues. In contrast, the proposed method outperformed all others, achieving a 2.26\% higher Dice score than the second-best Gated CRF method, demonstrating its superior generalization ability. This is because the theory of evidence provides a solid mathematical foundation for EDL, enabling the model to accurately quantify prediction uncertainty. The evidence loss function directed the model to minimize prediction errors while also minimizing the under- or overestimation of uncertainty. This robust learning process facilitated the model's ability to learn more generalized feature representations, rather than merely memorizing noise or specific patterns within the training data.

\begin{table*}[ht]
	\begin{center}
		\caption{Evaluation of model robustness on the BUSI Dataset. The symbol $\sigma$ represents the standard deviation of the Gaussian noise added to the image. The right subscript of the Dice coefficient denotes the magnitude of variation between noise-perturbed and original results, where a negative sign indicates performance degradation.}
		\label{tab:robustness_BUSI}
		\setlength{\tabcolsep}{4mm}{
			\begin{tabular}{ c | c  c | c  c }
				\toprule
				\multirow{2}{*}{Methods} & Dice (\%) $\uparrow$ & 95HD (pixel) $\downarrow$ & Dice (\%) $\uparrow$ & 95HD (pixel) $\downarrow$ \\
				\cmidrule(lr){2-5}
				& \multicolumn{2}{c|}{No noise} & \multicolumn{2}{c}{$\sigma=0.05$} \\
				\midrule
				U-Net + pCE & 62.89 $\pm$ 2.90 & 47.03 $\pm$ 5.77 & 60.32 $\pm$ 3.85\textsubscript{-2.57} & 49.41 $\pm$ 3.67 \\
				U-Net + Gated CRF & \textit{71.39 $\pm$ 2.66} & 35.90 $\pm$ 2.86 & \textit{69.71 $\pm$ 2.73}\textsubscript{-1.69} & 37.85 $\pm$ 1.99 \\
				U-Net + USTM & 61.99 $\pm$ 3.05 & 47.50 $\pm$ 5.16 & 60.41 $\pm$ 3.83\textsubscript{\textit{-1.58}} & 46.42 $\pm$ 4.96 \\
				DMPLS & 65.09 $\pm$ 3.44 & 36.27 $\pm$ 3.21 & 62.80 $\pm$ 3.91\textsubscript{-2.30} & 36.82 $\pm$ 6.49 \\
				ScribbleVC & 69.92 $\pm$ 3.87 & \textit{31.23 $\pm$ 5.67} & 67.84 $\pm$ 3.30\textsubscript{-2.08} & 34.24 $\pm$ 3.56 \\
				ScribFromer & 65.09 $\pm$ 3.44 & 36.27 $\pm$ 3.21 & 55.51 $\pm$ 3.09\textsubscript{-9.58} & 45.53 $\pm$ 2.10 \\
				DMSPS & \textit{71.75 $\pm$ 2.14} & 31.26 $\pm$ 3.40 & 69.53 $\pm$ 2.30\textsubscript{-2.22} & \textit{33.37 $\pm$ 3.66} \\
				Bayesian\_WSS & 68.42 $\pm$ 3.02 & 34.78 $\pm$ 8.13 & 66.46 $\pm$ 2.85\textsubscript{-1.97} & 34.13 $\pm$ 2.78 \\
				Ours & \textbf{74.07 $\pm$ 3.27} & \textbf{31.10 $\pm$ 2.73} & \textbf{72.92 $\pm$ 3.99}\textsubscript{\textbf{-1.15}} & \textbf{30.38 $\pm$ 2.47} \\
				\midrule
				& \multicolumn{2}{c|}{$\sigma=0.10$} & \multicolumn{2}{c}{$\sigma=0.15$} \\
				\midrule
				U-Net + pCE & 52.58 $\pm$ 4.15\textsubscript{-10.31} & 55.96 $\pm$ 8.79 & 38.71 $\pm$ 9.08\textsubscript{-24.18} & 66.86 $\pm$ 14.73 \\
				U-Net + Gated CRF & \textit{62.23 $\pm$ 4.26}\textsubscript{\textit{-9.16}} & 41.58 $\pm$ 5.98 & \textit{48.60 $\pm$ 10.58}\textsubscript{-22.79} & \textbf{41.87 $\pm$ 7.13} \\
				U-Net + USTM & 50.63 $\pm$ 5.89\textsubscript{-11.35} & 45.19 $\pm$ 8.31 & 31.89 $\pm$ 14.46\textsubscript{\textit{-30.10}} & \textit{47.87 $\pm$ 15.54} \\
				DMPLS & 45.96 $\pm$ 5.88\textsubscript{-19.13} & 54.66 $\pm$ 23.12 & 22.20 $\pm$ 6.39\textsubscript{-42.89} & 55.01 $\pm$ 27.41 \\
				ScribbleVC & 53.40 $\pm$ 10.40\textsubscript{-16.52} & 56.38 $\pm$ 17.52 & 35.21 $\pm$ 7.48\textsubscript{-34.71} & 87.06 $\pm$ 16.43 \\
				ScribFromer & 54.69 $\pm$ 3.64\textsubscript{-10.41} & 46.02 $\pm$ 1.84 & 43.35 $\pm$ 10.79\textsubscript{\textit{-21.74}} & 64.86 $\pm$ 25.09 \\
				DMSPS & 57.95 $\pm$ 4.21\textsubscript{\textit{-13.80}} & \textit{31.07 $\pm$ 4.25} & 33.04 $\pm$ 7.37\textsubscript{\textit{-38.71}} & 51.69 $\pm$ 15.05 \\
				Bayesian\_WSS & 57.61 $\pm$ 4.77\textsubscript{-10.81} & 35.74 $\pm$ 7.71 & 35.35 $\pm$ 11.09\textsubscript{-33.08} & 48.86 $\pm$ 23.56 \\
				Ours & \textbf{65.54 $\pm$ 3.76}\textsubscript{\textbf{-8.53}} & \textbf{39.21 $\pm$ 4.22} & \textbf{55.11 $\pm$ 2.40}\textsubscript{\textbf{-18.97}} & 51.55 $\pm$ 14.14 \\
				\bottomrule
		\end{tabular}}
	\end{center}
\end{table*}

\subsubsection{Robustness analysis}
When applying models in real-world clinical scenarios, they sometimes need to confront noisy data. Therefore, it is necessary to test the performance of the model on noisy data to validate its robustness. We applied Gaussian noise to degrade the image quality on both the BUSI dataset and the CardiacUDA dataset, simulating scenarios of low-quality data acquisition. We considered three distinct levels of Gaussian noise, characterized by standard deviations $\sigma$ of 0.05, 0.1, and 0.15, respectively, under the condition that image pixels had been normalized to a range of 0 to 1. As evident from the experimental results presented in Tables~\ref{tab:robustness_BUSI} and~\ref{tab:robustness_CardiacUDA}, when $\sigma=0.05$, the performance degradation observed for all methods was relatively small, whereas our method maintained optimal performance with the least degradation, demonstrated by a mere 0.31\% decrease in Dice score on the CardiacUDA dataset. When theta increased to 0.1, our method remained optimal with minimal performance degradation. However, under extreme conditions where theta reaches 0.15, while most methods experienced significant performance deterioration, our method still retained an advantage, demonstrating the best performance with the least degradation. Specifically, on the BUSI dataset, our method outperformed the worst-performing DMPLS Dice by a significant margin of 32.91\%, demonstrating a substantial advantage. On the CardiacUDA dataset, our performance degradation was minimal, at just 3.81\%. We observed significant disparities in model performance between the BUSI and CardiacUDA datasets, with models experiencing more pronounced performance degradation on BUSI as noise levels increase. We hypothesize that this could be attributed to the richer supervision information provided by the multi-class data. Given the larger number of classes, models may require learning more intricate features to represent the data, enabling them to better adapt and overcome the impact of noise. In summary, our method exhibited robustness due to the Dirichlet distribution's capability to model probabilities of probabilities, essentially the confidence distribution of prediction outcomes across different classes. This approach offered greater flexibility by incorporating evidence theory to quantify prediction uncertainty, enabling the model to distinguish whether uncertainty stems from data noise or limitations in its own knowledge. This explicit modeling of uncertainty aided the model in maintaining stability when confronted with complex or anomalous inputs.

\begin{figure*}[ht]
	\includegraphics[width=1\linewidth]{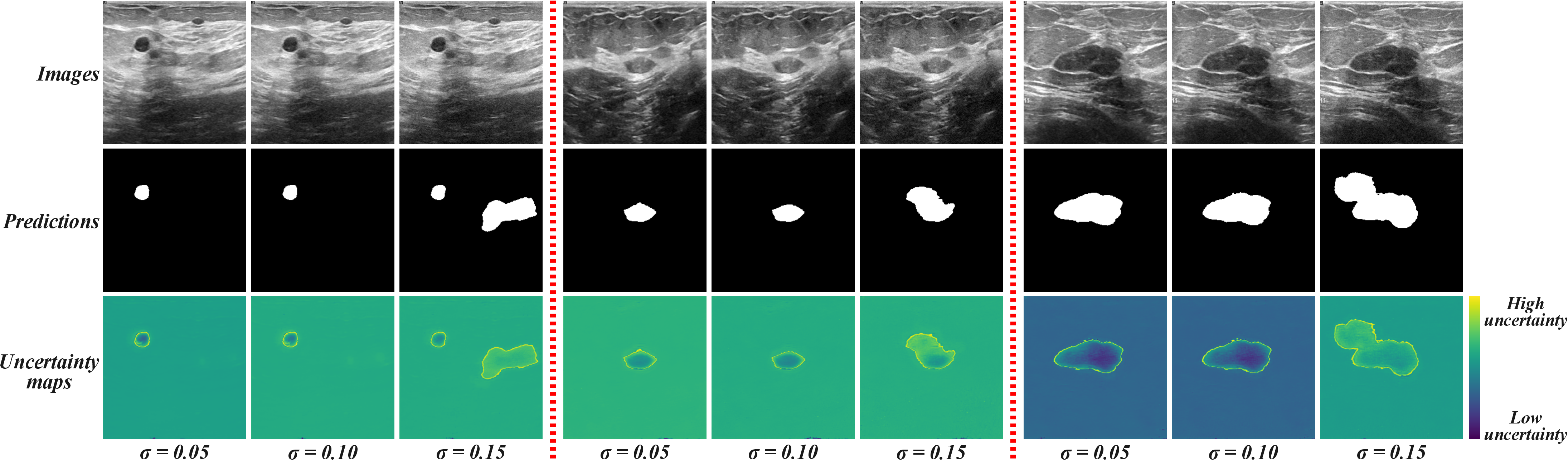}
	\caption{Visualization of uncertainty maps for the test results of our model on the BUSI dataset. The symbol $\sigma$ represents the standard deviation of the Gaussian noise added to the image. The figure presents three ultrasound image samples (separated by red dashed lines), each subjected to three distinct levels of Gaussian noise.}
	\label{fig:robust_BUSI}
\end{figure*}

\begin{figure*}[ht]
	\includegraphics[width=1\linewidth]{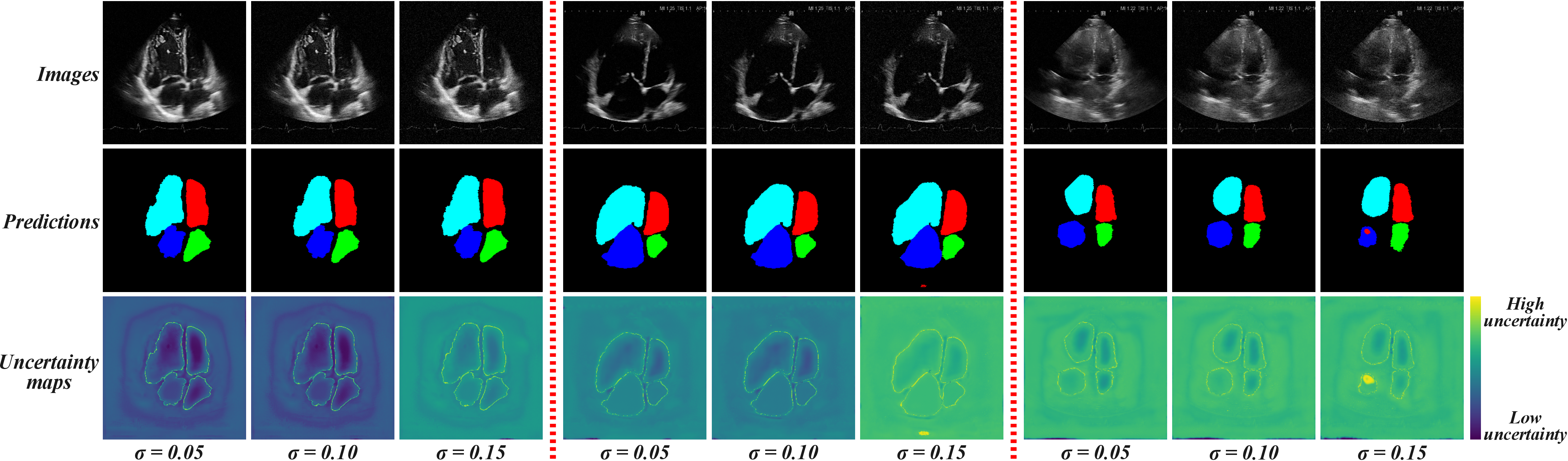}
	\caption{Visualization of uncertainty maps for the test results of our model on the CardiacUDA dataset. The symbol $\sigma$ represents the standard deviation of the Gaussian noise added to the image. The figure presents three ultrasound image samples (separated by red dashed lines), each subjected to three distinct levels of Gaussian noise.}
	\label{fig:robust_CardiacUDA}
\end{figure*}

Furthermore, we visualized the uncertainty maps, as shown in Figures~\ref{fig:robust_BUSI} and~\ref{fig:robust_CardiacUDA}. In most cases, the overall uncertainty increased with the increment of noise. The model's predictions tended to exhibit higher uncertainty at the edges of the target regions in individual images, while the internal regions of the targets generally showed lower uncertainty. As noise increased, some areas may be incorrectly predicted, but these regions typically manifested high uncertainty, observable from the figures. Notably, Figure~\ref{fig:robust_CardiacUDA} revealed the notable impact of scribble annotations on some image predictions, with low uncertainty regions emerging around scribble annotations in background areas. This observation validated the effectiveness of our uncertainty estimation, as predictions within annotated regions should be highly evidential.

\section{Discussion}
The purpose of WSL based on scribble annotations is to reduce annotation workload while maintaining good performance. Consistency regularization is generally based on the smoothness assumption. This strategy is rooted in the idea that, given a sample typically located in a high-density region, generating a perturbed sample from it and assuming both are in high-density regions, consistency regularization enforces their predictions to be consistent. This paper proposed utilizing two distinct architectures, CNN and Mamba (with differing feature extraction capabilities, namely global and local), to generate perturbed sample features and then enforce consistency in their predictions. However, considering ultrasound images and sparse scribble annotations, the segmentation of boundaries is often ambiguous. Directly discarding predictions at these decision boundaries can impair performance. Therefore, we proposed using EDL to estimate the prediction confidence of ambiguous pixels by the two models and then used high-confidence predictions in high-density regions to guide the optimization of low-confidence predictions at decision boundaries. We conducted extensive experiments on four ultrasound datasets, including common organs like the heart, thyroid, and breast, as well as both binary and multi-class segmentation problems. Both quantitative analysis and visualization results demonstrated the competitiveness of our method. Visualization results showed that our method handled boundaries better. Our method not only outperformed other methods in performance on multi-class segmentation tasks but also exhibited small performance differences between categories, unlike some methods that tended to segment easier categories better. Moreover, our method avoided serious misidentifications of categories such as heart chambers, which were common errors in other models. In binary segmentation tasks, our method can handle targets of different sizes and shapes without the over-segmentation or under-segmentation issues common in some models, partly due to the excellent global feature capture capability of the Mamba branch and the fusion of global and local information. Ablation experiments validated the effectiveness of all proposed components and strategies. By introducing advanced Mamba, which significantly reduced computational complexity compared to Transformers, we replaced the Mamba branch with other advanced models for comparison. Experimental results showed that it not only offered superior performance but also lower computational costs. Additionally, due to the introduction of EDL, the robustness of the model was enhanced. Therefore, we tested the robustness of the model, and experimental results indicate that our method has strong generalization ability across different datasets and noise adaptability.

\begin{table*}[ht]
	\begin{center}
		\caption{Evaluation of model robustness on the CardiacUDA Dataset. The symbol $\sigma$ represents the standard deviation of the Gaussian noise added to the image. The right subscript of the Dice coefficient denotes the magnitude of variation between noise-perturbed and original results, where a negative sign indicates performance degradation.}
		\label{tab:robustness_CardiacUDA}
		\setlength{\tabcolsep}{4mm}{
			\begin{tabular}{ c | c  c | c  c }
				\toprule
				\multirow{2}{*}{Methods} & Dice (\%) $\uparrow$ & 95HD (pixel) $\downarrow$ & Dice (\%) $\uparrow$ & 95HD (pixel) $\downarrow$ \\
				\cmidrule(lr){2-5}
				& \multicolumn{2}{c|}{No noise} & \multicolumn{2}{c}{$\sigma=0.05$} \\
				\midrule
				U-Net + pCE & 64.76 $\pm$ 8.18 & 37.59 $\pm$ 26.48 & 64.39 $\pm$ 1.22\textsubscript{\textit{-0.37}} & 31.50 $\pm$ 0.92 \\
				U-Net + Gated CRF & \textit{74.87 $\pm$ 2.03} & \textit{11.72 $\pm$ 4.43} & \textit{74.18 $\pm$ 0.49}\textsubscript{-0.69} & \textit{12.61 $\pm$ 0.93} \\
				U-Net + USTM & 64.82 $\pm$ 7.40 & 38.74 $\pm$ 24.70 & 63.67 $\pm$ 1.94\textsubscript{-1.15} & 35.05 $\pm$ 1.06 \\
				DMPLS & 66.10 $\pm$ 5.80 & 28.79 $\pm$ 19.19 & 65.68 $\pm$ 0.73\textsubscript{-0.42} & 28.38 $\pm$ 0.42 \\
				ScribbleVC & 71.45 $\pm$ 1.92 & 14.14 $\pm$ 4.12 & 70.98 $\pm$ 2.22\textsubscript{-0.47} & 17.00 $\pm$ 6.80 \\
				ScribFromer & 61.90 $\pm$ 4.96 & 29.41 $\pm$ 10.40 & 61.42 $\pm$ 1.33\textsubscript{-0.48} & 29.97 $\pm$ 4.81 \\
				DMSPS & 73.28 $\pm$ 2.20 & 12.59 $\pm$ 4.23 & 72.58 $\pm$ 1.27 \textsubscript{-0.70} & 12.97 $\pm$ 0.51 \\
				Bayesian\_WSS & 74.34 $\pm$ 1.63 & 12.08 $\pm$ 4.14 & 73.72 $\pm$ 1.10\textsubscript{-0.62} & 12.53 $\pm$ 0.69 \\
				Ours & \textbf{75.39 $\pm$ 1.60} & \textbf{11.60 $\pm$ 4.33} & \textbf{75.08 $\pm$ 0.65}\textsubscript{\textbf{-0.31}} & \textbf{12.07 $\pm$ 0.48} \\
				\midrule
				& \multicolumn{2}{c|}{$\sigma=0.10$} & \multicolumn{2}{c}{$\sigma=0.15$} \\
				\midrule
				U-Net + pCE & 56.96 $\pm$ 7.65\textsubscript{-7.8} & 31.36 $\pm$ 2.30 & 40.95 $\pm$ 13.83\textsubscript{-23.81} & 31.34 $\pm$ 4.51 \\
				U-Net + Gated CRF & \textit{73.06 $\pm$ 0.56}\textsubscript{\textit{-1.81}} & \textit{13.32 $\pm$ 0.76} & \textit{69.71 $\pm$ 2.09}\textsubscript{\textit{-5.16}} & \textit{17.85 $\pm$ 5.98} \\
				U-Net + USTM & 51.78 $\pm$ 4.59\textsubscript{-13.04} & 29.29 $\pm$ 2.27 & 30.40 $\pm$ 7.69\textsubscript{-34.42} & 24.89 $\pm$ 4.18 \\
				DMPLS & 59.55 $\pm$ 2.82\textsubscript{-6.55} & 30.54 $\pm$ 2.28 & 41.87 $\pm$ 6.24\textsubscript{-24.23} & 32.96 $\pm$ 6.61 \\
				ScribbleVC & 68.98 $\pm$ 5.19\textsubscript{-2.47} & 18.28 $\pm$ 7.97 & 65.64 $\pm$ 6.90\textsubscript{-5.81} & 21.62 $\pm$ 8.45 \\
				ScribFromer & 59.38 $\pm$ 2.55\textsubscript{-2.52} & 32.03 $\pm$ 6.46 & 55.97 $\pm$ 6.62\textsubscript{-5.93} & 33.83 $\pm$ 7.19 \\
				DMSPS & 70.99 $\pm$ 1.59\textsubscript{-2.29} & 14.65 $\pm$ 1.68 & 66.54 $\pm$ 4.37\textsubscript{-6.74} & 20.41 $\pm$ 6.60 \\
				Bayesian\_WSS & 72.28 $\pm$ 1.46\textsubscript{-2.06} & 14.15 $\pm$ 1.32 & 67.83 $\pm$ 3.46\textsubscript{-6.51} & 19.93 $\pm$ 4.28 \\
				Ours & \textbf{74.01 $\pm$ 0.72}\textsubscript{\textbf{-1.38}} & \textbf{13.04 $\pm$ 1.02} & \textbf{71.58 $\pm$ 1.31}\textsubscript{\textbf{-3.81}} & \textbf{15.89 $\pm$ 1.86} \\
				\bottomrule
		\end{tabular}}
	\end{center}
\end{table*}

While our study advances scribble-based weakly supervised ultrasound segmentation, several limitations warrant discussion. First, our current framework focuses exclusively on sparse scribble annotations within ultrasound imaging; future work should explore unified methods that synergize sparse and dense supervision paradigms while extending applicability to cross-modal scenarios (e.g., pancreatic lesion segmentation in CT/MRI) to validate generalizability beyond ultrasound-specific challenges. Second, although our method operates effectively without pre-training, emerging ultrasound foundation models could enhance annotation efficiency through hybrid architectures that integrate generalized representations with our evidence-guided strategy, preserving computational efficiency while improving boundary awareness in low-data regimes. Third, the mutual exclusivity assumption in our evidential framework may limit performance in regions with ambiguous acoustic boundaries. Future extensions could incorporate anatomical prior knowledge (e.g., cardiac chamber adjacency) via spatially-aware Dirichlet priors or hierarchical evidence accumulation to resolve such ambiguities. These directions aim to broaden clinical applicability while addressing fundamental challenges in ultrasound segmentation.

\section{Conclusion}
We have proposed a scribble-based WSL approach for ultrasound image segmentation, effectively reducing the annotation cost. The CNN-Mamba dual-branch networks that have been introduced effectively capture and fuse both local and global features while mitigating computational complexity. Moreover, the EDL-based EGC strategy that has been proposed refines the model's prediction of edge regions, preserving and leveraging low-confidence pixels to enhance the stability of target edge prediction. Extensive experiments conducted on four public ultrasound image datasets have demonstrated that our method has exhibited superior performance in both binary and multi-class tasks, outperforming existing WSL approaches in terms of performance, generalization, and robustness. These accomplishments validate the effectiveness of our approach and its potential for clinical applications.

\section{Acknowledgments}
This work was supported by the National Natural Science Foundation of China (Grant No. 62071285), the Eastern Scholars Program from Shanghai Municipal Education Commission, the Fundamental Research Funds for the Central Universities (Grant No. YG2025QNA07), the Pudong New Area Science and Technology Development Fund (Grant No. PKJ2025-Y04), and the Shanghai Technical Service Center of Science and Engineering Computing, Shanghai University.

\bibliography{references}
\end{document}